%% file: preprint.tex
\documentclass[letterpaper,doc,12pt,floatsintext,longtable]{apa6}

\newcommand{\papertitle}{Estimating Optimal Dynamic Treatment Regimes for Personalized Education: \break A Tutorial and Applications with Machine Learning}   % include \\[0.35em] for a line break
\newcommand{\shortpapertitle}{ODTR for Personalized Education}
\newcommand{\authornames}{Chenguang Pan$^{1,*}$, Yuxuan Li$^{1}$, Thomas Harris$^{2}$, and Youmi Suk$^{1}$}
\newcommand{\institution}{{\small $^{1}$Department of Human Development, Teachers College, Columbia University \\
$^{2}$Department of Mathematics, Science and Technology, Teachers College, Columbia University\normalsize}}
\newcommand{\authoremails}{\{cp3280, yl5655, thh2117, ysuk\}@tc.columbia.edu}
\newcommand{\paperdate}{September 22, 2026}
\newcommand{\journalname}{Zeitschrift f\"ur Psychologie}
\newcommand*{\journaldoiurl}{}
\newcommand{\correspondingauthor}{Chenguang Pan}
\newcommand{\correspondingemail}{cp3280@tc.columbia.edu}
\newcommand{\abstracttext}{Optimal dynamic treatment regimes (ODTRs) are increasingly used to generate sequences of personalized recommendations that maximize a final outcome of interest. Although ODTRs are widely used in biostatistics and precision medicine, their applications in education remain limited, especially for longitudinal decision-making. This study translates ODTR methods for single- and multi-stage designs into individualized high school math course recommendations, providing step-by-step guidelines and demonstrations with data from the High School Longitudinal Study of 2009 (HSLS:09). We use Targeted Maximum Likelihood Estimation with ensemble machine learning to estimate ODTRs that maximize students' math achievement and college enrollment, imposing feasibility constraints based on domain knowledge and propensity-score thresholds to ensure recommendations are practically implementable. Results show that unconstrained regimes yield higher estimated values but infeasible recommendations, while feasible regimes produce realistic pathways with stronger empirical and expert-informed support. Finally, we discuss practical considerations for designing ODTRs in education.}
\newcommand{\keywordlist}{optimal policy learning, optimal dynamic treatment regimes, targeted maximum likelihood estimation, personalized education, causal machine learning}
\usepackage[american]{babel}
\usepackage[utf8]{inputenc}
\usepackage{amsmath}
\usepackage{amssymb}
\usepackage{bm}
\usepackage{graphicx}
\usepackage{xcolor}
\usepackage{colortbl}
\usepackage{caption}
\usepackage{subcaption}
\usepackage{setspace}
\usepackage{threeparttable}
\usepackage{endnotes}
\usepackage{xspace}
\usepackage{afterpage}
\usepackage{lscape}
\usepackage{longtable} % apa6 loads longtable only in man mode; the variable list and glossary need it
\usepackage{arydshln}  % must precede hyperref, or links to longtables (Tables A1, S1) lose their targets
\usepackage[colorlinks=true, allcolors=black]{hyperref}
\hypersetup{pdftitle={Estimating Optimal Dynamic Treatment Regimes for Personalized Education: A Tutorial and Applications with Machine Learning},
            pdfauthor={Chenguang Pan, Yuxuan Li, Thomas Harris, Youmi Suk}}
\usepackage[normalem]{ulem}
\usepackage{csquotes}
\usepackage{dsfont}
\usepackage{booktabs}
\usepackage{multirow}
\usepackage{tabularx}
\usepackage{array}
\usepackage[noend]{algpseudocode}
\usepackage{algorithm}
\usepackage{listings}
\usepackage{bbm}

\usepackage{geometry}
\usepackage{tikz}
\usepackage{pgf}
\usetikzlibrary{
    matrix,
    positioning,
    calc,
    arrows,
    shapes.arrows,
    shapes.geometric,
    shapes.multipart,
    decorations.pathreplacing,
    decorations.pathmorphing,
    shapes,
    decorations,
    arrows.meta,
    fit
}

\definecolor{confcol}{HTML}{E6F1FB}\definecolor{confdraw}{HTML}{185FA5} % confounders
\definecolor{predcol}{HTML}{E1F5EE}\definecolor{preddraw}{HTML}{0F6E56} % predictors
\definecolor{neutcol}{HTML}{F1EFE8}\definecolor{neutdraw}{HTML}{5F5E5A} % A, Y, noise
\definecolor{modcol}{HTML}{BA7517}                            % effect-modifier ring

\tikzset{
  var/.style   = {circle, draw, minimum size=10mm, inner sep=1pt, font=\small,
                  line width=0.5pt},
  conf/.style  = {var, fill=confcol, draw=confdraw},
  pred/.style  = {var, fill=predcol, draw=preddraw},
  focal/.style = {var, fill=neutcol, draw=neutdraw, line width=0.9pt},
  noise/.style = {rectangle, rounded corners=3pt, draw=neutdraw, dashed,
                  fill=neutcol, font=\small, align=center, inner sep=5pt},
  mbadge/.style= {circle, fill=modcol, text=white, font=\bfseries\tiny,
                  inner sep=0.6pt},
  mod/.style   = {draw=modcol, line width=1pt, double, double distance=1.2pt,
                  label={[mbadge, label distance=-1.6mm]45:M}},
  cedge/.style = {-{Stealth[length=2mm]}, draw=confdraw, line width=0.5pt},
  pedge/.style = {-{Stealth[length=2mm]}, draw=preddraw, line width=0.5pt},
  tedge/.style = {-{Stealth[length=2mm]}, draw=neutdraw, line width=1pt},
  lkey/.style  = {circle, minimum size=4mm, inner sep=0pt, line width=0.5pt},
}

\usepackage{listings}
\graphicspath{{./}{figures/}}

\usepackage[style=apa,sortcites=true,sorting=nyt,backend=biber]{biblatex}
\DeclareLanguageMapping{american}{american-apa}
\newcommand{\PreserveBackslash}[1]{\let\temp=\\#1\let\\=\temp}
\newcolumntype{C}[1]{>{\PreserveBackslash\centering}p{#1}}
\newcolumntype{L}[1]{>{\PreserveBackslash\raggedright}p{#1}}
\newcolumntype{R}[1]{>{\PreserveBackslash\raggedleft}p{#1}}

\DeclareMathOperator*{\argmaxA}{arg\,max}

\newcommand\independent{\protect\mathpalette{\protect\independenT}{\perp}}
\def\independenT#1#2{\mathrel{\rlap{$#1#2$}\mkern2mu{#1#2}}}

\newcommand{\esmsec}[1]{\ref{#1}}
\newcommand{\esm}[1]{Section~\esmsec{#1} of ESM~1}

\renewcommand{\arraystretch}{1.1}
\title{\fontsize{14}{17}\selectfont \papertitle}
\shorttitle{\shortpapertitle}
\author{\fontsize{13.5pt}{13pt}\selectfont \authornames \vspace{0.3em}}
\affiliation{\institution \\[0.3em]
\authoremails
\\[0.5em]
\paperdate \\[0.3em]
{\small Accepted for publication in \textit{\journalname} (in press)}}

\begin{document}

\maketitle

% Corresponding-author footnote (symbol marker, not a number)
\begingroup
\renewcommand{\thefootnote}{\fnsymbol{footnote}}
\footnotetext[1]{\textbf{Correspondence:} \correspondingauthor, Email: \correspondingemail\par
\textbf{Data Availability:} This study uses restricted-use data from the High School Longitudinal Study of 2009 (HSLS:09). The license terms do not allow the authors to share these data, but qualified researchers can apply for a license through the Standard Application Process described at \url{https://nces.ed.gov/statprog/instruct.asp}. Annotated code and a synthetic dataset are available at \url{https://github.com/cgpan/feasibleODTR}.\par
% Required by Hogrefe for posted accepted manuscripts; the "at <DOI>" part appears once \journaldoiurl is set
\textbf{Version notice:} This version of the article may not completely replicate the final authoritative version published in \textit{\journalname}\ifx\journaldoiurl\empty\else{} at \expandafter\url\expandafter{\journaldoiurl}\fi. It is not the version of record and is therefore not suitable for citation.}
\endgroup

% ----------------------------- ARXIV ABSTRACT --------------------------------
\vspace{-1em}
\begin{center}
\textbf{Abstract}
\end{center}

\noindent
\begin{minipage}{\textwidth}
\abstracttext

\vspace{0.75em}

\end{minipage}

\vspace{1.5em}

% Number down to subsubsection level
\setcounter{secnumdepth}{3}

% =============================================================================
% BODY -- EDIT FROM HERE
% =============================================================================

\section{Introduction}
Personalized decision-making has been adopted across medicine, computer science, education, and psychology. In education, it aims to tailor learning pathways to students' needs, but school practice remains largely heuristic and relies on teachers' and counselors' judgments of ability and interest \parencite{walkington2020appraising, dumont2023promise}. Precision medicine has moved further toward data-driven approaches based on optimal treatment regimes (OTRs) and their multi-stage extension, optimal dynamic treatment regimes (ODTRs), which use patients' characteristics and history to find heterogeneous treatment effects and convert them into a sequence of decision rules maximizing a final outcome \parencite{tsiatis2019dynamic, murphy2007developing}.

To illustrate this in an educational context, consider high school mathematics in the United States, where students typically follow a common course-taking sequence such as Algebra I, Geometry, Algebra II, and advanced courses \parencite{burkam2003mathematics, dalton2007advanced}. Such a one-size-fits-all plan ignores meaningful variation in optimal educational pathways depending on students' prior performance, motivation, and interests. In the terminology we adopt below, this conventional sequence is a static treatment regime, in which the entire series of courses is fixed before the first decision is made and is never revised in light of intermediate evaluations. A school could assign each entering ninth grader a distinct four-year plan based on baseline information alone, and that plan would remain static as long as no intervening information could alter it. An ODTR instead specifies a decision rule at each grade that maps a student's accumulated history (e.g., intermediate grades, completed courses, and evolving interests) into the next course, so that ODTRs recommend the pathway maximizing final educational outcomes such as end-of-high-school math achievement or college enrollment. Because the static class is nested within the dynamic class, the ODTR is never worse than an optimal static regime in the population under the identification assumptions stated in Section~\ref{sec:assumptions}. Although static regimes retain practical appeal for their simplicity and for preserving the peer groupings that a dynamic regime may disrupt, we focus on ODTR methods and return to static regimes in Section~\ref{sec:design}.

Recently, researchers have coupled the ODTR method with machine learning (ML) for flexible function estimation \parencite{chernozhukov2018double, xiong2022artificial}, but applications in education remain scarce \parencite{suk2023designing}, especially in longitudinal settings where decisions unfold over several years. To increase the accessibility of ODTRs, the overarching goal of this paper is to provide applied researchers with enough of the logic, notation, and implementation detail to build and evaluate an ODTR themselves, using mathematics course-taking as a worked illustration. We make three contributions. First, we provide a reproducible workflow for single- and multi-stage regimes using Targeted Maximum Likelihood Estimation \parencite[TMLE,][]{vanderlaan2011targeted} with ensemble ML, which decomposes packaged software into explicit fitting steps and supplies annotated code and a synthetic dataset so readers can run every step efficiently. For brevity, we call the multi-stage approach ODTR-TMLE and its single-stage version OTR-TMLE. Second, we provide demonstrations and discussions on applying OTR-TMLE and ODTR-TMLE in a nationally representative dataset, the restricted version of the High School Longitudinal Study 2009 \parencite[HSLS:09;][]{ingels2011hsls}, with feasibility constraints that take both school policy and empirical support into consideration. Lastly, we provide guidelines for applying ODTR methods in education. Table~\ref{tab:glossary} in the Appendix gives working definitions of the technical terms used throughout the paper.

\section{A Running Example}
To illustrate the method and the technical concepts beneath it, we use the HSLS:09 dataset as a running example and construct mathematics course recommendation models in both a single-decision and a multiple-decision setting to optimize math learning outcomes. The analytic sample has 12,640 students in 800 schools after data cleaning (sample sizes are rounded to the nearest 10 to comply with Institute of Education Sciences disclosure rules). 

In the single-decision setting, the decision point occurs in Grade 9. Because the raw transcript GPA cannot distinguish course rigor, such as the same letter grade A in Algebra I and Pre-calculus, we use the weighted math GPA at the end of Grade 9 as the outcome variable, assigning each course level a rigor weight. We provide details about how we construct the weighted GPA in Section~\esmsec{app:GPA} of Electronic Supplementary Material 1 (ESM~1). The treatment is the highest-level math course a student completes in that academic year (termed ``math pipeline'' in the HSLS:09 documentation), which has seven ordered categories: ``No Math'' (coded as 1), ``Other Math'' (2), ``Basic Math'' (3), ``Algebra I'' (4), ``Geometry'' (5), ``Algebra II'' (6), and ``Advanced Math'' (7).

\begin{table}[htbp]
  \centering
  \begin{threeparttable}
    
    % Caption: Table number is automatic
    \caption{Course Treatment Levels by Grade}
    \label{tab:treatments}

    % Table content
    % 8 columns: 4 pairs of (Level, Course)
    \begin{tabular*}{\linewidth}{@{\extracolsep{\fill}}cccccccc@{}}
      \toprule
      
      % Spanning Headers for the Groups (9, 10, 11, 12)
      \multicolumn{2}{c}{Grade 9} & \multicolumn{2}{c}{Grade 10} & \multicolumn{2}{c}{Grade 11} & \multicolumn{2}{c}{Grade 12} \\
      
      % Rules grouping the headers (lr trims the line ends for separation)
      \cmidrule(lr){1-2} \cmidrule(lr){3-4} \cmidrule(lr){5-6} \cmidrule(lr){7-8}
      
      % Sub-headers
      Level & Course & Level & Course & Level & Course & Level & Course \\
      \midrule
      
      % Data Rows
      1 & NM & 1 & NM & 1 & NM & 1 & NM \\
      2 & OM & 2 & OM & 2 & OM & 2 & OM \\
      3 & BM & 3 & BM & 3 & BM & 3 & BM \\
      4 & A1 & 4 & A1 & 4 & A1 & 4 & A1 \\
      5 & GM & 5 & GM & 5 & GM & 5 & GM \\
      6 & A2 & 6 & A2 & 6 & A2 & 6 & A2 \\
      7 & AM & 7 & AB & 7 & PS & 7 & PS \\
        &    & 8 & PC & 8 & AM & 8 & AM \\
        &    & 9 & CL & 9 & AB & 9 & AB \\
        &    &   &    & 10 & PC & 10 & PC \\
        &    &   &    & 11 & CL & 11 & CL \\
        &    &   &    & 12 & AC & 12 & AC \\
      
      \bottomrule
    \end{tabular*}
       \footnotesize 
    %\begin{spacing}{1.0}  % adjust the bewteen line spaces.
    \noindent
    \textbf{NOTE}: NM = No Math, OM = Other Math, BM = Basic Math, A1 = Algebra 1, GM = Geometry, A2 = Algebra 2, PS = Probability \& Statistics, AM = Advanced Math, AB = AP/IB Math, PC = Pre-calculus, CL = Calculus, AC = AP Calculus. We aggregate sparse categories into a single composite level to mitigate data sparsity. \newline 
    \textbf{SOURCE}: Department of Education, National Center for Education Statistics, High School Longitudinal Study of 2009 (HSLS:09).
  \end{threeparttable}
\end{table}

In the multi-stage setting, decision points occur at Grades 9, 10, 11, and 12. We estimate ODTRs for two outcomes. The math achievement model maximizes math performance at the end of high school, while the college enrollment model aims to increase the probability of enrollment in a Bachelor's degree program. The outcome for the first model is the cumulative weighted math GPA at the end of high school, which pools all mathematics courses on the Grade 9--12 transcript so that the outcome summarizes achievement over the full course-taking sequence rather than the final year alone. The grade-specific weighted math GPAs serve as time-varying indicators representing students' historical math progress (see \esm{app:GPA} for detailed discussion). The second model uses a binary indicator for enrollment in a college for a Bachelor's degree as the outcome. The two models share the same set of treatment variables that are the grade-specific math pipelines, including the 7-level treatment defined in the single-decision model. Table~\ref{tab:treatments} summarizes treatment definitions across grades.

Both settings condition on a similar covariate set, covering demographics, socioeconomic status and parental education, pre-high-school mathematics preparation, mathematics and science motivation, and school characteristics like the mathematics-teacher-to-student ratio, sector, locale, and climate. Missing values are imputed by predictive mean matching \parencite{white2011multiple}. In the multi-stage setting, the historical covariate set at each decision point adds whatever has been realized since the previous one, so from Grade 10 onward, it includes the prior year's mathematics weighted GPA and course. Because each stage regression conditions on those entire accumulated historical characteristics, estimation requires a wide format, i.e., one row per student, with stage-specific columns. 

A moderator is a covariate that drives the heterogeneity and makes the learned regime personalized, as it allows rules to tailor recommendations to individual student profiles rather than relying on population-average effects. Within the covariate set, we also pre-specify candidate moderators based primarily on domain knowledge, such as prior math achievement and math identity, which often relate to course placement and differential benefit from the advanced coursework \parencite[]{mceachin2020heterogeneous,simpkins2006math}. We remark that moderators can also be chosen in a data-adaptive way, especially when domain knowledge is limited. The moderators can be set to full covariate history and rely on flexible algorithms with built-in regularization (e.g., penalty-based variable selection in LASSO) to represent effect modification through their functional form. \esm{app:var_list} lists the full variable set.

\section{Estimands and Identification Assumptions}

\subsection{Notation and Estimands}

Following \textcite{tsiatis2019dynamic}, with $K$ decision points the observed data for individual $i$ are $O_i=(X_i^{(1)},A_i^{(1)},\dots, X_i^{(K)}, A_i^{(K)},Y_i)$. In our dataset, $K=4$, indexing Grades 9 through 12. Baseline covariates $X_i^{(1)}$ are eighth-grade preparation, socioeconomic status, initial mathematics identity, and school characteristics. $X_i^{(k)}$ for $k \geq 2$ accrue between decisions, such as the mathematics GPA earned in Grade 9. The treatment $A_i^{(k)} \in \mathcal{A}^{(k)}$ is the mathematics course taken at stage $k$, and $Y_i$ the final outcome. The history collects everything known when a decision is made:
\begin{align*}
H_i^{(1)} = X_i^{(1)}, \qquad
H_i^{(k)} = \left(X_i^{(1)},A_i^{(1)},\dots,A_i^{(k-1)}, X_i^{(k)}\right), \quad k = 2, \dots, K.
\end{align*}
$H^{(k)}$ does not contain the current treatment $A^{(k)}$. For example, a rising eleventh grader's history holds baseline characteristics, the Grade 9 and 10 courses, and the grades earned in them, but not the Grade 11 course, which is the decision at hand. 

We also define $\overline{X}_i^{(k)}=(X_i^{(1)},X_i^{(2)},\cdots, X_i^{(k)})$ and $\overline{A}_i^{(k)}=(A_i^{(1)},A_i^{(2)},\cdots, A_i^{(k)})$, so that $H_i^{(k)}=(\overline{X}_i^{(k)}, \overline{A}_i^{(k-1)})$. Let $\overline{\mathcal A}^{(k)}=\mathcal A^{(1)} \times \cdots \times \mathcal A^{(k)}$ denote the set of all treatment histories through decision point $k$, and let $\overline{a}^{(k)}=(a^{(1)},\dots,a^{(k)}) \in \overline{\mathcal A}^{(k)}$ denote a generic treatment history through stage $k$. For notational simplicity, let $\overline{A}_i=\overline{A}_i^{(K)}$, $\overline{\mathcal A}=\overline{\mathcal A}^{(K)}$, and $\overline{a}=\overline{a}^{(K)}$ denote the full observed treatment history, its support, and a generic element of that support. A decision rule $d^{(k)}:\mathcal{H}^{(k)} \rightarrow \mathcal{A}^{(k)}$ maps a student's history $H^{(k)}$ to a math course in the available set ${\mathcal A}^{(k)}$, and a dynamic treatment regime $d=(d^{(1)},\dots,d^{(K)})$ collects the rules across decision points, which is a complete four-year math course-taking plan in our data. Let $\mathcal{D}$ denote the class of all such regimes, so that $d \in \mathcal{D}$. 

Let $Y_i^*(d)$ denote the outcome realized if at each stage individual $i$ received $A_i^{(k)}=d^{(k)}(H_i^{*(k)})$, where $H_i^{*(k)}$ is the history that would unfold under $d$. Under a dynamic regime, that history is itself a potential outcome \parencite{rubin1974estimating}, since what the rule recommends in Grade 11 depends on the Grade 10 GPA. The value $V(d)=\mathbb{E}\{Y^*(d)\}$ is the average outcome if the whole population followed the regime, and an ODTR $d^{\mathrm{opt}}$ is any regime in $\mathcal{D}$ maximizing it. We remark that not every option is available to every student in a real school setting. For example, a student who has not passed Pre-calculus cannot take Calculus, and a small, rural school may not offer AP Calculus. Let $\mathcal A_{\mathrm{feas}}^{(k)}(h^{(k)}) \subseteq \mathcal A^{(k)}$ denote the nonempty set of options feasible at stage $k$ for history $h^{(k)}$, and let $\mathcal D_{\mathrm{feas}}$ denote the class of feasible regimes, those whose rule at every stage recommends only options feasible for the current history,
\begin{equation}\label{eq:feasible_class}
\mathcal D_{\mathrm{feas}}=\bigl\{d\in\mathcal D:\ d^{(k)}(h^{(k)})\in\mathcal A_{\mathrm{feas}}^{(k)}(h^{(k)}) \text{ for all } h^{(k)}\in\mathcal H^{(k)} \text{ and } k=1,\dots,K\bigr\}.
\end{equation}
The estimand becomes
\begin{equation}\label{eq:feasible_optimal_regime}
d_{\mathrm{feas}}^{\mathrm{opt}} = \argmaxA_{d \in \mathcal{D}_{\mathrm{feas}}} \ V(d),
\end{equation}
the optimal feasible regime a school could implement. Since $\mathcal D_{\mathrm{feas}} \subseteq \mathcal D$, necessarily $V(d_{\mathrm{feas}}^{\mathrm{opt}}) \leq V(d^{\mathrm{opt}})$. Any loss of value reflects real-world constraints rather than a defect of the method, so the feasibility-value trade-off is worth reporting. We discuss this trade-off in Section~\ref{sec:OTR-demo}.

We construct the feasible sets $\mathcal A_{\mathrm{feas}}^{(k)}(h^{(k)})$ from three kinds of restrictions with different justifications. Structural constraints exclude options that are genuinely impossible, such as unmet prerequisites or courses not offered, so that $P(A^{(k)}=a \mid H^{(k)}=h^{(k)})$ is truly zero. Normative constraints exclude options ruled out by local policy or professional judgment rather than empirical possibility. Under these constraints, certain choices observed in the historical data are intentionally removed from the feasible decision set. For example, after fulfilling core graduation requirements, some Grade 12 students historically choose lower-level math or forgo math entirely. By removing these suboptimal courses from the available set, normative constraints aim to recommend advanced coursework and enhance college enrollment chances. Statistical constraints exclude options that are possible but rarely observed for a student's given history, so the data cannot support estimating their effects, which remedies practical positivity violations discussed in Section~\ref{sec:assumptions} and is not a claim that an option is educationally infeasible. We implement the first two through \textit{a priori} rules and the third through a propensity threshold \parencite{van2007realistic, pan2026learning}, and an option is feasible for a history only if none of the three excludes it. Conflating the third with the first is a serious interpretive error, and Section~\ref{sec:design} discusses the equity risk it creates.

\subsection{Assumptions and the Plausibility in Practice}\label{sec:assumptions}
To identify the value of a feasible treatment regime from observed data, we rely on four causal assumptions \parencite{tsiatis2019dynamic, rubin1986comment, imbens2015causal}. These assumptions allow causal quantities defined in terms of potential outcomes to be expressed using the observed data distribution. We provide explanations and honest discussion of their plausibility in real-world educational data.

\begin{itemize}
    \item[(A1)] Consistency and no hidden variation:
    $$
    X_i^{(k)}
    = X_i^{*(k)}\!\left(\overline{A}_i^{(k-1)}\right)
    = \sum_{\overline{a}^{(k-1)} \in \overline{\mathcal{A}}^{(k-1)}}
    X_i^{*(k)}\!\left(\overline{a}^{(k-1)}\right)\,
    \mathbbm{I}\!\left(\overline{A}_i^{(k-1)} = \overline{a}^{(k-1)}\right),
    \quad k = 2,\dots,K,
    $$
    \begin{equation*}
    Y_i
    = Y_i^*\!\left(\overline{A}_i\right)
    = \sum_{\overline{a} \in \overline{\mathcal{A}}}
    Y_i^*\!\left(\overline{a}\right)\,
    \mathbbm{I}\!\left(\overline{A}_i = \overline{a}\right).
    \end{equation*}

    \item[(A2)] No interference: One individual's potential outcomes do not depend on others' assignments.
    
    \item[(A3)] Sequential Unconfoundedness:
    $$
    A_i^{(k)} \independent W_i^* \mid H_i^{(k)}, \qquad k = 1, \cdots, K,
    $$
    where $W_i^*$ is the set of all possible potential outcomes,
    \begin{align*}
    W_i^*
    = \bigl\{
    &
    X_i^{*(2)}(a^{(1)}),
    X_i^{*(3)}(\overline{a}^{(2)}),
    \cdots,
    X_i^{*(K)}(\overline{a}^{(K-1)}),
    Y_i^*(\overline{a}),\\
    &\text{for }
    a^{(1)} \in \mathcal{A}^{(1)},
    \overline{a}^{(2)} \in \overline{\mathcal{A}}^{(2)},
    \cdots,
    \overline{a}^{(K-1)} \in \overline{\mathcal{A}}^{(K-1)},
    \overline{a} \in \overline{\mathcal{A}}
    \bigr\}.
    \end{align*}
    
    \item[(A4)] Restricted positivity over feasible options: For each stage $k=1,\dots,K$, any history $h^{(k)}$ with $P\!\left(H^{(k)}=h^{(k)}\right)>0$, and every feasible option $a \in \mathcal A_{\mathrm{feas}}^{(k)}(h^{(k)})$,
    $$
    P\!\left(A^{(k)}=a \mid H^{(k)}=h^{(k)}\right) > 0.
    $$
    
\end{itemize} 

Assumption (A1) has two components that are usually collapsed. Consistency narrowly requires that being assigned a course by a rule is functionally identical to enrolling by choice, which is questionable in an educational setting. A student placed into Algebra I by an algorithm while friends are placed into Geometry may respond through resentful demoralization or reduced effort, and one granted unexpected access to acceleration may experience a motivational boost that choosing the course would not have produced. The second component, no hidden variation, requires that each treatment option name a single well-defined exposure, and in HSLS:09 it almost certainly fails. Many schools offer Algebra I and Algebra II at several levels of pace and rigor---honors, regular, stretched two-year sequences---all coded identically. Because the treatment is the highest-level course completed, a student enrolled in more than one mathematics course in the same year is represented only by the highest one, and a student repeating Algebra I after failing receives the same code as a first-time enrollee. Additionally, the course content varies across the 800 schools sampled. Each label is therefore a coarsening of several distinct exposures. This estimate reflects a blended mix of environments, so that it cannot reliably predict the impact of a decision that changes those conditions. For instance, if an assignment rule recommends many students into Algebra II, newly enrolled students will likely face different tracks, class sizes, and teachers than the historical students who generated the original estimate.

Assumption (A2) is close to untenable in a school. Classrooms are shared environments in which peers affect learning through study groups, norms, competition, and the pace an instructor can sustain, so implementing an ODTR at scale would reassign a substantial share of students and alter the composition of every classroom. We do not attempt to relax this assumption. Instead, the honest reading of the estimates in educational data is as individual-level advisory quantities, which describe what one additional student might expect while holding the system fixed.

Assumption (A3) generalizes ``no unmeasured confounding'' to the longitudinal setting and requires that the course taken at each grade is as good as randomly assigned among students with the same history. Its conditioning set is history-specific and grows over time, which makes (A3) more plausible than a baseline-only version, since placement is driven heavily by the previous year's grade and course, both conditioned on at every stage. We therefore include rich student- and school-level covariates as confounders as described in \esm{app:var_list} and expect the observed history to absorb a substantial share of the confounding association, helped by multi-item scales for mathematics self-efficacy, interest, and identity that are unusual among large-scale transcript studies. However, educational data may not cover all possible confounders. For example, HSLS:09 does not measure mathematics anxiety, which predicts both the avoidance of advanced coursework and performance conditional on enrollment. Our assessment is that the observed covariates likely remove a substantial portion, but not all, of the confounding association. A formal sensitivity analysis is the most important extension we do not undertake but provide discussion in Section~\ref{sec:limits}. Finally, (A4) requires positive probability only for feasible options \parencite{van2007realistic, pan2026learning}. With 12 mathematics options at Grade 11, global positivity would be hopeless, whereas restricted positivity is attainable.

Assumptions (A1) and (A2) are unlikely to hold in these data in our judgment, and (A3) is partially met. We therefore do not claim that the estimated regimes identify causally optimal placements. What the estimates support is that the workflow runs on real longitudinal educational data at realistic scale, reveals what unconstrained and constrained ODTR produce when the option set is large and unevenly populated, and quantifies a trade-off between the feasibility and overall outcome that any applied user will confront. We accordingly describe the results below as model-implied recommendations rather than optimal placements, and read estimated values as comparisons among fitted regimes under stated assumptions rather than as forecasts. Researchers wanting stronger claims will need treatment definitions with less hidden variation, a design that confronts interference, and sensitivity analysis.

\section{Estimation for ODTR} \label{sec: est_ODTR}
%%%%%%%%%%%%%%%%%%%%%%%%%%%%%%%%%%%%%%%%%%%%%%%%%%%%%%%%%%%%%
The core difficulty in a multi-stage ODTR task is that early decisions may have delayed effects. A Grade 9 math course that maximizes the first-year math GPA may not guarantee a maximal outcome at the end of high school. Backward induction solves this issue by starting at the final decision stage \parencite[]{tsiatis2019dynamic}. Because no decision follows the last decision, the optimal Grade 12 rule is unambiguous, which is to take the one with the highest expected final outcome (math achievement or college enrollment) among feasible options given everything that has happened. Once that rule is fixed, we know what a student would obtain from Grade 12 onward for any Grade 11 history, and that continuation outcome becomes the target for the Grade 11 decision. The argument repeats backward to Grade 9. Formally, at the final decision point, define
\begin{equation}\label{eq:Q}
Q^{(K)}(h^{(K)},a^{(K)}) = \mathbb E\!\left\{Y \mid H^{(K)}=h^{(K)},\, A^{(K)}=a^{(K)}\right\},
\end{equation}
read as ``the expected final outcome for a student with this past record who now takes this course.'' The letter $Q$ denotes ``quality,'' following the reinforcement-learning literature \parencite[]{murphy_experimental_2005, watkins_technical_1992} and it represents nonparametric regression. The optimal stage-$K$ value and rule are
\begin{equation}\label{eq:VK}
\begin{aligned}
V^{(K)}(h^{(K)})&=\max_{a^{(K)} \in \mathcal A_{\mathrm{feas}}^{(K)}(h^{(K)})} Q^{(K)}(h^{(K)},a^{(K)}),\\
d_{\mathrm{feas}}^{(K)\mathrm{opt}}(h^{(K)})&\in\mathop{\text{argmax}}_{a^{(K)} \in \mathcal A_{\mathrm{feas}}^{(K)}(h^{(K)})} Q^{(K)}(h^{(K)},a^{(K)}).
\end{aligned}
\end{equation}
The $\text{max}$ operator returns the best expected outcome attainable once the optimal rule is assumed, while the $\text{argmax}$ returns the course attaining it. If a Grade 12 student's feasible set is \{Calculus, Probability and Statistics, AP Calculus\} with predicted cumulative GPAs of 3.10, 3.05, and 2.90, then their $V^{(4)}=3.10$ and $d_{\mathrm{feas}}^{(4)\mathrm{opt}}= \text{Calculus}$. The recommendation uses the $\text{argmax}$, while the recursion uses the $\max$, because 3.10 is what the Grade 11 decision should position this student to achieve. For earlier stages, the target is that continuation value $V$ rather than the final outcome $Y$,
\begin{equation}\label{eq:Qk}
Q^{(k)}(h^{(k)},a^{(k)}) = \mathbb E\!\left\{V^{(k+1)}\!\left(H^{(k+1)}\right) \mid H^{(k)}=h^{(k)},\, A^{(k)}=a^{(k)}\right\},
\end{equation}
with $V^{(k)}$ and $d^{(k)\mathrm{opt}}$ defined analogously. At $k=3$ this asks: for a student with this record through Grade 10 who now takes this Grade 11 course, what is the expected value of the best available Grade 12 continuation? The regression target is the future (Grade 12) value, not the Grade 11 value, which is precisely how delayed effects enter. This recursive fitting is known as \emph{Q-learning} \parencite{murphy_experimental_2005}. Alternative methods include G-estimation \parencite{robins2004optimal} and value search \parencite{zhang2012estimating}.

Rather than estimating $Q^{(k)}(h^{(k)},a^{(k)})$ separately for each option and ranking the fitted values, in practice, it is more stable to estimate contrasts between options directly since the errors common to all options cancel in a difference. For a fixed reference option $a_1^{(k)}$, the blip function $B^{(k)}_{j,1}(h^{(k)}) = Q^{(k)}(h^{(k)}, a_j^{(k)}) - Q^{(k)}(h^{(k)}, a_1^{(k)})$ is the expected gain from taking $a_j^{(k)}$ instead of the reference, given optimal treatment thereafter. The term ``blip'' comes from structural nested mean models \parencite{robins2004optimal}, where a blip is a brief departure from the reference option at a single decision point, with the student's earlier record unchanged and all later decisions following the optimal rules. The blip function asks questions like how much higher this student's expected end-of-high-school GPA would be under Geometry rather than Basic Math. We note that the blip function is a stage-specific Conditional Average Treatment Effect (CATE), which directly captures treatment heterogeneity across history states $h^{(k)}$. Optimal decision rules can be built on the blips by recommending the feasible option with the largest blip, where the reference itself has a blip of zero.

Because $Q$ is unknown, blips are estimated. For each option $a\in\mathcal A^{(K)}$, we form a doubly robust pseudo-outcome by augmented inverse probability weighting \parencite{Robins1994Estimation}:
\begin{equation}\label{eq:aipw_K_1}
D^{(K)}_{a}(O)
=
\frac{\mathbbm{1}\{A^{(K)}=a\}}{\hat e^{(K)}(a\mid H^{(K)})}
\left\{Y-\hat Q^{(K)}(H^{(K)},a)\right\}
+\hat Q^{(K)}(H^{(K)},a),
\end{equation}
which combines the model's prediction of what a student would achieve under that option, $\hat Q^{(K)}(H^{(K)},a)$, with a correction that applies only for students who actually received it, $Y-\hat Q^{(K)}(H^{(K)},a)$, weighted inversely by how unlikely the option was for them, $\hat e^{(K)}(a\mid H^{(K)})$. Students who took an improbable option therefore carry more weight, as they are the only observed examples of what happens under that choice for similar students, which is the source of both the double robustness and the variance problems noted below. Differences between the pseudo-outcomes for two options are the blip pseudo-outcomes, whose conditional mean given $H^{(k)}$ is the blip function. Therefore, regressing them on the history turns a causal contrast into an ordinary regression problem.

In causal inference literature, nuisance functions refer to propensity score $e^{(k)}(a \mid h) = P(A^{(k)}=a \mid H^{(k)}=h)$ and the outcome regression $Q^{(k)}$, since they are needed to reach the target but not of substantive interest, and misspecifying them biases everything downstream. In practice, instead of relying on a single algorithm, we estimate the nuisance functions using Super Learner \parencite{vanderlaan2007super}, an ensemble ML method with a built-in cross-validation (CV) process. The analyst specifies a library of candidate algorithms, such as a generalized linear model, LASSO, random forest, and gradient boosting, and the data are split into $S$ blocks, or folds, with each candidate trained on $S-1$ and scored on the remainder, and the candidates are combined into a weighted average minimizing cross-validated error, with the guarantee that the ensemble performs asymptotically at least as well as its best member \parencite{luedtke2016super}. Library composition therefore matters more than any single algorithmic choice, since only linear models cannot represent effect modification operating through interactions or complex non-linear relations, and practitioners may save time by avoiding selecting a single algorithm and tuning the model. CV extends the logic to the target parameter, estimating nuisances on training folds and evaluating only on held-out folds, so the same observations never both learn a rule and evaluate it. Without CV, flexible fits carry an overfitting bias into the value estimate, and intervals are too narrow \parencite{chernozhukov2018double}. 

Averaging the fitted regression under the learned rule to get the estimate (i.e., the estimated value) is not sufficient for valid inference, because ML estimators are tuned to minimize global prediction error rather than optimizing the estimate of a specific target parameter. They leave a first-order bias in the quantity of interest and carry no valid standard errors. TMLE therefore applies a small targeted update to the initial regression, designed to remove that bias in the value parameter specifically and driven by a term built from the estimated propensity scores, historically called the clever covariate \parencite{vanderlaan2011targeted, van2015target}. The result is a plug-in estimator respecting parameter bounds, doubly robust in that it is consistent if either the outcome regression or the propensity model is, and asymptotically linear with a known influence function, so it yields valid intervals despite the use of ML. Our inferential target is therefore $V(\hat d_{\mathrm{feas}})$, the value of the sample-learned rule treated as fixed on held-out data, so our 95\% confidence intervals cover the value of the rule this procedure produced, rather than the value of the unknown population-optimal feasible regime. 

To see how personalization operates, we walk through two illustrative ninth graders. Student A enters with weak pre-high-school preparation and low mathematics identity, and student B enters with strong preparation and high identity. The feasible sets can differ from the start, because an option such as Algebra II may fall below the propensity threshold given A's history while remaining feasible for B. Suppose the potential outcomes under Algebra I and Geometry are 2.9 and 2.6 for A but 3.0 and 3.3 for B, the rule then recommends Algebra I for A and Geometry for B, because the same course carries different expected outcomes for different profiles. Personalization is also dynamic. If B later completes Pre-calculus with a strong Grade 11 record, AP Calculus may carry the highest predicted continuation value at Grade 12, whereas after a weak Grade 11 the same student may be placed into Probability and Statistics, so the recommendation tracks both who the student is at baseline and what has happened since.

The performance of ODTR-TMLE in practice depends on several conditions. The causal work is done by the identification assumptions discussed in Section~\ref{sec:assumptions}, so unmeasured confounding biases the value estimate regardless of the estimator, and extreme propensity scores directly inflate the inverse-propensity weights, which is the practical motivation for the propensity threshold in the feasibility constraints. Sample size matters per option and per stage rather than in total, because every outcome and treatment model is learned from students with similar histories who took each option, and rarely chosen options at late stages are precisely where estimates are least reliable. A large covariate set does not require manual pre-screening, since penalized learners in the Super Learner library, such as LASSO, perform automated variable selection and tree-based learners perform implicit selection, although irrelevant covariates still add variance and computation time.

We provide a comprehensive description of the technical details of ODTR-TMLE in \esm{app:tech_details}. To facilitate the applications, we also provide commented source code, demonstrations using synthetic data, and other replication materials at the first author’s GitHub repository: \url{https://github.com/cgpan/feasibleODTR}.

\section{Single-Decision OTR for Personalized Education} \label{sec:OTR-demo}
In this section, we provide a step-by-step tutorial of building a math course recommendation model for ninth graders to maximize expected math GPA at the end of Grade 9. We estimate OTRs under three settings: (i) a standard OTR without feasibility constraints; (ii) a feasible OTR with \textit{a priori} constraints and a propensity-score cutoff of 0.03; and (iii) a feasible OTR with \textit{a priori} constraints and a propensity-score cutoff of 0.05. \textcite{pan2026learning} suggests a recommended cutoff range of 0.01 to 0.08 for multicategorical treatments. Accordingly, we use 0.03 and 0.05 to represent relatively lenient and relatively strict trimming of empirically unsupported treatment options.

We impose an \textit{a priori} constraint based on students' pre-high school math achievement to restrict recommendations that are inconsistent with foundational readiness. This score is a weighted composite derived from students' self-reported scores of the highest math course completed before high school. We set two cutoffs. The first cutoff is 3.6, approximately equivalent to earning a B in Math 8 or Pre-algebra. Students whose pre-high school math achievement score is below 3.6 are not recommended to take courses above Algebra I. The second cutoff is 4.2, approximately equivalent to earning a B in Algebra I. Students whose scores exceed 4.2 are not recommended to take courses below Algebra I. This \textit{a priori} constraint is normative rather than structural and serves as a practical proxy for student readiness and placement rules. Therefore, it does not attempt to reproduce formal prerequisite policies that are not observed in HSLS:09.

\subsection{Steps}
Since this application uses a single decision point, we write the observed data as $O=(X,A,Y)$ and omit stage superscripts for simplicity. We use ten-fold cross-fitting throughout. In each fold, we estimate nuisance functions on the training split and generate predictions on the held-out split. Unless otherwise stated, all quantities below are based on cross-fitted predictions.

\paragraph{Step 1: Estimate nuisance functions.}
We estimate the outcome regression $Q(x,a)=\mathbb E(Y\mid X=x,A=a)$ and the propensity score $e(a\mid x)=P(A=a\mid X=x)$ using Super Learner. To efficiently handle various possible data structures, we employ algorithms from parametric models (e.g., Generalized Linear Model, LASSO regression) to flexible models (e.g., random forest, XGBoost), and rely on the built-in meta-learning process to adaptively weight the results from each candidate. This step returns cross-fitted predictions $\hat Q(X_i,a)$ for each treatment level $a\in\mathcal A$ and $\hat e(a\mid X_i)$ for each individual $i$.

\paragraph{Step 2: Construct pseudo-outcomes and estimate blip functions.}
We adopt and simplify Equation~\eqref{eq:aipw_K_1} in this step. For each treatment level $a\in\mathcal A$, we compute the doubly robust pseudo-outcome
$$
D_a(O)=
\frac{\mathbbm{1}\{A=a\}}{\hat e(a\mid X)}\{Y-\hat Q(X,a)\}+\hat Q(X,a).
$$
We select a reference level $a_1$ and define blip pseudo-outcomes
$$
\tilde D_{j,1}(O)=D_{a_j}(O)-D_{a_1}(O),\qquad j=2,\dots,J.
$$
We then estimate the conditional blip functions $B_{j,1}(x)=\mathbb E\{\tilde D_{j,1}(O)\mid X=x\}$ using Super Learner, which yields fitted contrasts $\hat B_{j,1}(x)$ for decision making. The candidate ML algorithms for blip estimation should include both parametric and data-adaptive models to capture broader statistical relationships. In addition to comparing each pseudo-outcome with the first option to construct the blips, we note that one can also use the average or weighted-average pseudo-outcome across all treatment levels as the reference.

\paragraph{Step 3: Apply feasibility constraints and learn a feasible rule.}
The feasibility constraints step primarily uses the estimated $\hat B_{j,1}(x)$ and the full-set-predicted propensity-score matrix $\hat e(a\mid x)$. 

Statistical models may yield infeasible recommendations if they fail to account for institutional or contextual constraints. Besides, global positivity is easily violated with multicategorical treatments. We therefore restrict decisions to a feasible set $\mathcal A_{\mathrm{feas}}(x)\subseteq\mathcal A$ defined by domain knowledge and a propensity-score threshold. In implementation, we set $\hat B_{j,1}(x)=-\infty$ for options $a_j\notin\mathcal A_{\mathrm{feas}}(x)$ so that maximization occurs over feasible options only.

The learned feasible rule selects the feasible option with the largest estimated blip, where the reference itself has a blip of zero:
$$
\hat{d}_{\mathrm{feas}}(x)=\mathop{\mathrm{argmax}}_{a_j\in \mathcal A_{\mathrm{feas}}(x)}\hat B_{j,1}(x),\qquad \hat B_{1,1}(x)\equiv 0.
$$

\paragraph{Step 4: Estimate the value via TMLE}
After OTR learning, we estimate the value of the learned rule, $V(\hat{d}_{\mathrm{feas}})=\mathbb E\{Y^*(\hat {d}_{\mathrm{feas}})\},$ with cross-validated TMLE. In this step, the learned rule is treated as fixed on each validation split. Following the TMLE steps discussed in \esm{app:tech_details}, we estimate the updated value, the asymptotic variance, and construct a Wald-type confidence interval for the updated value. 

We conduct the data analysis in R \parencite[version 4.3.0;][]{R2025} using the \texttt{tmle3} \parencite[][]{jeremy2021tmle3}, \texttt{tmle3mopttx} \parencite[][]{malenica2022tmle3mopttx}, and \texttt{sl3} \parencite[][]{Jeremy2023sl3} packages. See \esm{app:R_app} for a brief introduction to the R application.

\subsection{Results}

\begin{figure}[!h]
    \centering
    \includegraphics[width=1\linewidth]{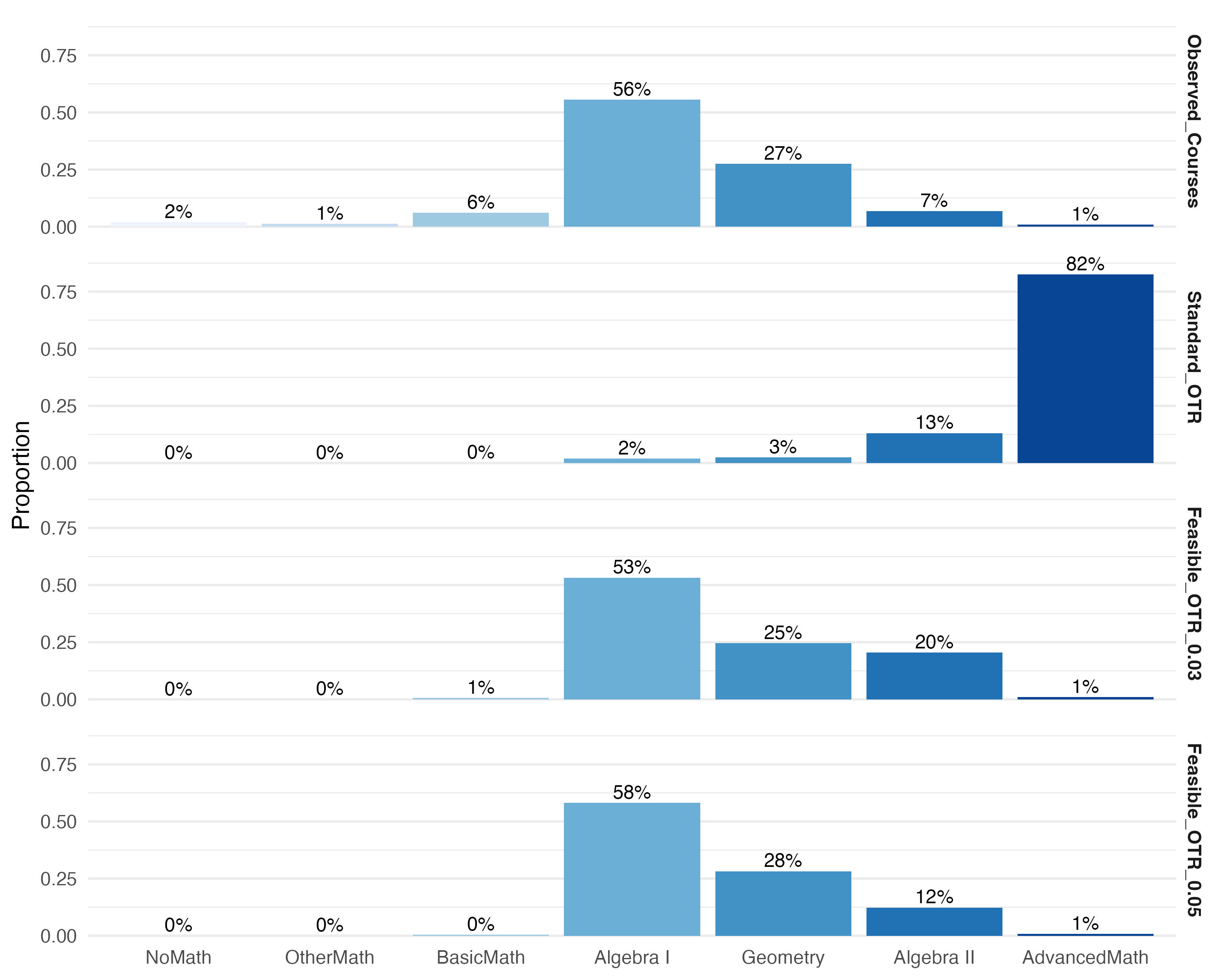}
    \caption{Comparison of math course placements under observed enrollment, a standard OTR, and feasible OTRs with \textit{a priori} constraints and propensity-score cutoffs of 0.03 and 0.05. Numbers shown as 0\% indicate estimated proportions below 0.05\% or categories with no predicted recommendations under the OTRs. In accordance with disclosure risk requirements, we do not report any results based on an unweighted $N<5$. \\ 
    SOURCE: Department of Education, National Center for Education Statistics, High School Longitudinal Study of 2009 (HSLS:09).}
    \label{fig:single_prop}
\end{figure}

Figure~\ref{fig:single_prop} compares the math course placements under (i) the observed course-taking patterns, (ii) standard OTR without any feasibility constraints, (iii) feasible OTR with \textit{a priori} constraints and a propensity-score cutoff at 0.03, and (iv) feasible OTR with \textit{a priori} constraints and a propensity-score cutoff at 0.05. In the observed data (top panel), Algebra I and Geometry are the most common choices and account for 55.5\% and 27.5\% of students, which reflects the typical early courses for high school math study. Under the standard OTR, 82\% of students are recommended to take the highest-level math course, as shown in the second panel. These recommended course-taking plans attain the largest expected grade-9 math GPA (i.e., the value under OTR) of 3.260 with an SE of 0.025, which is notably higher than the observed mean of 2.609. However, they are unlikely to be feasible in the real school setting, since this standard OTR model places most ninth graders directly into the Advanced Math by skipping the mandatory courses (i.e., Algebra I--Geometry--Algebra II). This unlikely pattern motivates the use of feasibility constraints to balance estimated utility with practical implementability.

The third panel shows that the feasible OTR under a relatively lenient propensity-score cutoff (0.03) retains Algebra I, Geometry, and Algebra II as dominant courses,  with the proportion of Algebra II increasing from 7\% in the observed data to 20\%. The corresponding estimated value is 2.747 (SE = 0.0288). Under the more stringent cutoff (0.05; bottom panel), recommendations shift toward more conservative placements, with increased allocations to Algebra I and Geometry and fewer recommendations to Algebra II. The point estimate decreases slightly to 2.746 (SE = 0.0290). By comparing the values from the standard OTR model with those from the feasible OTR models, we highlight a trade-off between estimated learning gains and feasibility restrictions, as stricter constraints reduce the estimated value in our results. As discussed in Section~\ref{sec:assumptions}, these values are model-implied comparisons among fitted regimes under the stated assumptions rather than forecasts of what adopting the recommendations would deliver. 

Practitioners can flexibly adjust the propensity-score cutoff and tailor the \textit{a priori} constraints to local policies and resource limits. When modifying these constraints, practitioners should account for the utility--feasibility trade-off observed here, where stricter feasibility restrictions can reduce the estimated value. 

\section{Multiple-Decision OTR for Personalized Education}\label{sec:ODTR-demo}
In this section, we provide step-by-step instructions for estimating ODTRs in a longitudinal setting with four decision points, corresponding to Grades 9 to 12. We construct models for two outcomes. The math achievement model aims to maximize the cumulative math GPA at the end of high school. The college enrollment model aims to increase the probability of enrollment in a degree-seeking college.

As in the single-decision setting, the ODTR analysis involves two components: (a) ODTR estimation; (b) value updating under the learned ODTRs \parencite[]{van2014target, van2015target}. We implement ODTR estimation via backward recursion from Grade 12 to Grade 9, consistent with the longitudinal decision structure described in Section~\ref{sec: est_ODTR} and \esm{app:tech_details}. Based on the cutoff range recommended by \textcite{pan2026learning}, we apply grade-specific propensity-score cutoffs to exclude empirically unlikely options, with $c_9=0.05$, $c_{10}=0.04$, $c_{11}=0.03$, and $c_{12}=0.03$. We also impose \textit{a priori} feasibility constraints based on prerequisites and course-taking policies. First, recommendations enforce Pre-calculus as a prerequisite for Calculus or AP Calculus. Second, because the Advanced Math category encompasses multiple distinct courses (e.g., Analytic Geometry, Abstract Algebra, and Linear Algebra), we allow this category to appear in multiple decision points in the recommended pathway.

\subsection{Steps}

This application uses $K=4$ decision points that correspond to Grades 9--12. We index stages by $k=1,\dots,4$, where $k=1$ denotes Grade 9 and $k=4$ denotes Grade 12. We partition the longitudinal data into ten folds and keep the same fold assignment across all stages. Unless stated otherwise, all nuisance models and targeting steps use cross-fitting. \esm{app:R_app} provides a brief introduction to R packages used for the ODTR task.

\paragraph{Step 1: Learn a feasible ODTR by backward induction.}
We begin at Grade 12. Let $Y^{(4)}=Y$ denote the final outcome, which is the cumulative weighted math GPA at the end of high school. Given the Grade 12 history $H^{(4)}$, we estimate the outcome regression $Q^{(4)}(h^{(4)},a^{(4)})=\mathbb E\!\left(Y^{(4)} \mid H^{(4)}=h^{(4)},A^{(4)}=a^{(4)}\right).$ Unlike the single-decision setting, the treatment models $\hat e^{(k)}(a^{(k)}\mid h^{(k)})=P\!\left(A^{(k)}=a^{(k)} \mid H^{(k)}=h^{(k)}\right), \quad k=1,\dots,4,$ must all be estimated together in Step~1 because the \textit{a priori} feasibility constraints at a later stage may depend on whether earlier-stage options are feasible. For example, Calculus in Grade 12 is feasible only if the Grade 11 course history in $h^{(4)}$ satisfies the Pre-calculus prerequisite, which in turn requires that the propensity score for taking Pre-calculus exceeds a pre-specified threshold. Otherwise, Calculus is treated as infeasible. We represent this structure through a history-dependent feasible set $\mathcal A^{(k)}_{\mathrm{feas}}(h^{(k)})\subseteq \mathcal A^{(k)}$.

Using $\hat Q^{(4)}$ and $\hat e^{(4)}$, we apply the single-decision workflow to obtain a feasible optimal rule $\hat d^{(4)}_{\mathrm{feas}}(h^{(4)})$ and an estimated stage-4 outcome $\hat V^{(4)}(h^{(4)})=\hat Q^{(4)}(h^{(4)},\hat d^{(4)}_{\mathrm{feas}}(h^{(4)}))$. For $k=3,2,1$, we define the pseudo-outcome $Y^{(k)}=\hat V^{(k+1)}(H^{(k+1)})$ and fit $Q^{(k)}(h^{(k)},a^{(k)})=\mathbb E(Y^{(k)}\mid H^{(k)}=h^{(k)},A^{(k)}=a^{(k)})$. We then learn $\hat d^{(k)}_{\mathrm{feas}}(h^{(k)})$ and set $\hat V^{(k)}(h^{(k)})=\hat Q^{(k)}(h^{(k)},\hat d^{(k)}_{\mathrm{feas}}(h^{(k)}))$. This recursion yields the learned feasible ODTR $\hat d_{\mathrm{feas}}=(\hat d^{(1)}_{\mathrm{feas}},\hat d^{(2)}_{\mathrm{feas}},\hat d^{(3)}_{\mathrm{feas}},\hat d^{(4)}_{\mathrm{feas}})$.

\paragraph{Step 2: Estimate the value of the learned ODTR via longitudinal TMLE}
As discussed in Section~\ref{sec: est_ODTR} and \esm{app:tech_details}, we estimate the value of $\hat d_{\mathrm{feas}}$ with longitudinal cross-validated TMLE by treating the fold-specific learned feasible rule as fixed on each validation split. Accordingly, the inferential target in this application is the mean outcome (i.e., the value) under the learned feasible ODTR, averaged across sample splits, rather than direct inference for the unknown population-optimal feasible regime.

We start at stage $k=4$. On the validation split, we construct the clever covariate by plugging $\hat d_{\mathrm{feas}}=(\hat d^{(1)}_{\mathrm{feas}},\hat d^{(2)}_{\mathrm{feas}},\hat d^{(3)}_{\mathrm{feas}},\hat d^{(4)}_{\mathrm{feas}})$ and $\hat e =(\hat e^{(1)},\hat e^{(2)},\hat e^{(3)},\hat e^{(4)} )$ and then update the initial regression $\hat Q^{(4)}_\mathrm{init}$ to $\hat Q^{(4)}_\mathrm{upd}$. We set the pseudo-outcome for stage-$3$ to $Y^{(3)}=\hat V^{(4)}_{\mathrm{upd}}(H^{(4)})$, where
$\hat V^{(4)}_{\mathrm{upd}}(h^{(4)})=\hat Q^{(4)}_\mathrm{upd}(h^{(4)},\hat d^{(4)}_{\mathrm{feas}}(h^{(4)}))$.

For $k=3,2,1$, we repeat the same procedure. We fit the initial regression $\hat Q^{(k)}_\mathrm{init}(h^{(k)},a^{(k)})=\mathbb E(Y^{(k)}\mid H^{(k)}=h^{(k)},A^{(k)}=a^{(k)})$ with Super Learner, construct the stage-specific clever covariate, and apply the same targeting step to obtain $\hat Q^{(k)}_\mathrm{upd}$. The final value estimate equals the empirical mean of updated Grade 9 value under the learned regime, $\hat V(\hat d_{\mathrm{feas}})=\frac{1}{n}\sum_{i=1}^n \hat V_{\mathrm{upd}}^{(1)}(X_i^{(1)}).$

Lastly, we compute the EIF, the empirical variance, and a Wald-type $95\%$ confidence interval. Because the estimator targets the mean outcome under the fold-specific learned feasible rules, the resulting inference pertains to this data-adaptive target parameter.

\paragraph{Step 3: ODTR quality evaluation.}
To assess the plausibility of recommendations, we define the empirical support rate as the proportion of students whose recommended course sequence appears in the observed course-taking patterns. Let $\hat{\overline{a}}_i$ denote the course sequence that $\hat{d}_{\mathrm{feas}}$ recommends to student $i$, and let $\mathcal S_{\mathrm{obs}}$ denote the set of distinct observed sequences. The support rate is
\begin{equation}\label{eq: quality_check}
\frac{1}{n}\sum_{i=1}^{n}\mathbbm{1}\!\left(\hat{\overline{a}}_i\in \mathcal S_{\mathrm{obs}}\right).
\end{equation}

In addition to the empirical support rate, practitioners may also like to check the alignment between the recommendations and the promising decisions based on the domain knowledge. For demonstration purposes, we define an expert-informed set of math pathways based on students' math and STEM identity, denoted by $\mathcal S_{\mathrm{expert}}$ (see \esm{app:theory-driven} for construction details). Replacing $\mathcal S_{\mathrm{obs}}$ with $\mathcal S_{\mathrm{expert}}$ in Equation~\eqref{eq: quality_check} yields an expert-alignment rate that quantifies concordance between model recommendations and expert expectations. The expert-alignment rate is optional and should be customized based on the research or application purpose. Both metrics assess recommendation plausibility. They do not by themselves establish causal optimality.

\subsection{Results}
After estimating the ODTRs, we evaluate the quality of recommended course-taking sequences and use the standard (unconstrained) ODTR models as a baseline. The math achievement model under the feasibility constraints yields an empirical support rate of 92.7\%, compared with 53.7\% for its unconstrained counterpart. For college enrollment, the constrained model attains 94.1\%, compared with 56.2\% for the standard model. As for the expert-alignment rates regarding the goal of math achievement, the observed course-taking patterns have a rate of 28.9\%, whereas the feasible ODTR increases this proportion to 43.9\%. For college enrollment, the corresponding proportion increases from 23.5\% in the observed data to 38.7\% under the feasible model. In contrast, standard ODTR rarely matches the expert-informed pathways, with proportions below 1\% for both outcomes. These improvements indicate that feasibility constraints improve alignment with both empirically observed patterns and domain-informed expectations.

\begin{figure}[!h]
    \centering
    \includegraphics[width=1\linewidth]{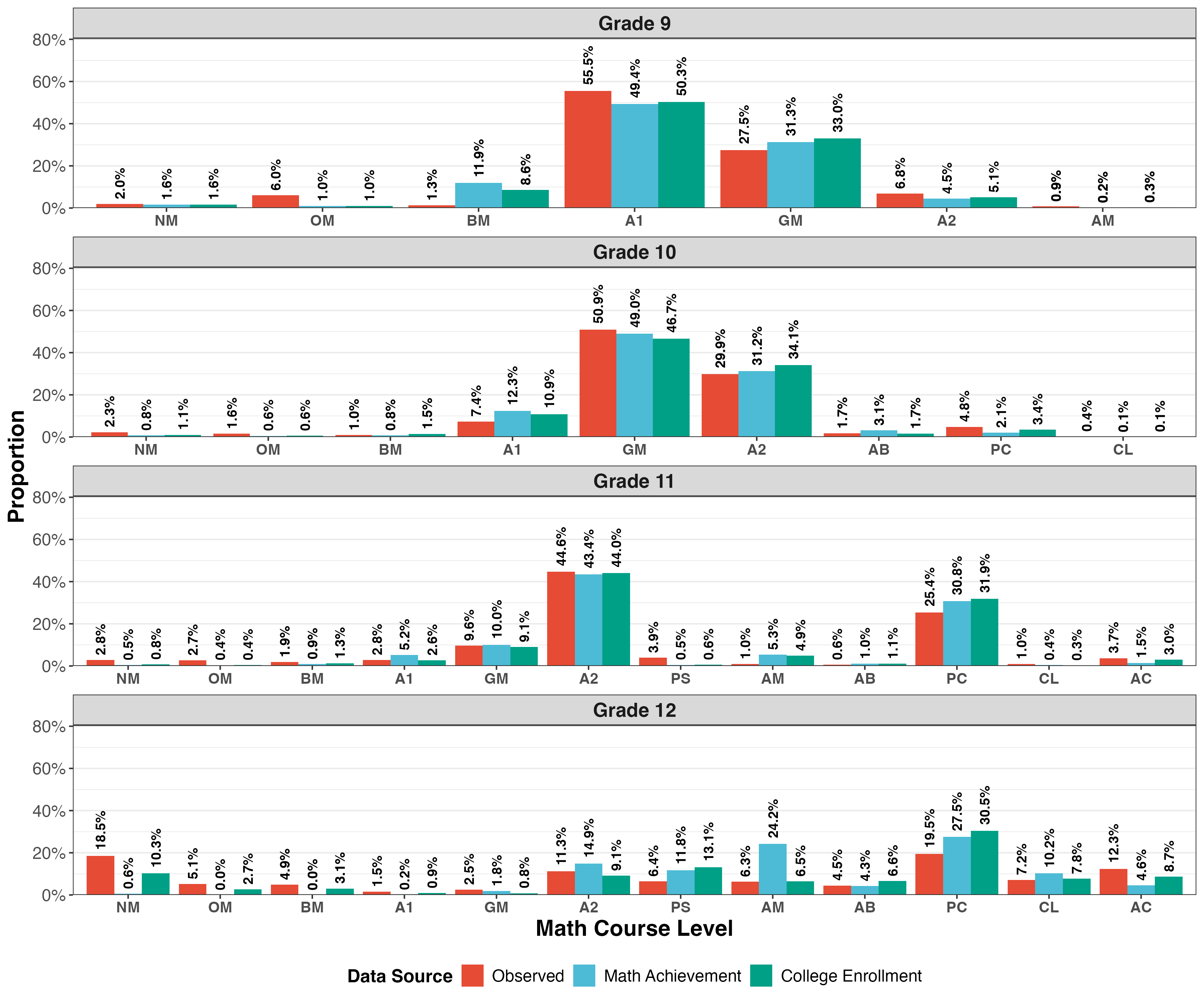}
    \caption{Comparison of math course proportions across four high school years: observed enrollment vs. feasible ODTR models for math achievement and college enrollment. Numbers shown as 0\% indicate estimated proportions below 0.05\% or categories with no predicted recommendations under the ODTRs. In accordance with disclosure risk requirements, we do not report any results based on an unweighted $N<5$.\\ 
    SOURCE: Department of Education, National Center for Education Statistics, High School Longitudinal Study of 2009 (HSLS:09).}
    \label{fig:multi_prop}
\end{figure} 

Figure~\ref{fig:multi_prop} compares marginal course allocations across grades under the observed data, the feasible math achievement model, and the feasible college enrollment model. Both models recommend slightly fewer students to the most common courses and more students to the second-most common courses in Grades 9--11. For example, Algebra I is the dominant observed Grade 9 course (55.5\%), whereas the two ODTRs recommend Algebra I for 49.4\% and 50.3\% of students. The ODTRs recommend Geometry for 31.3\% and 33.0\% of students, compared with 27.5\% observed. Both models shift a subset of students toward more advanced coursework when the estimated benefit is positive, and the recommendation is feasible. A key difference between the two models is that the college enrollment model recommends more calculus-related courses in Grades 11 and 12, which are important milestones for college preparation. 

The math achievement model attains an estimated value of 2.826 (SE = 0.043) under the feasible ODTRs, which exceeds the observed mean cumulative weighted math GPA of 2.770. The college enrollment model attains an estimated probability of 0.392 (SE = 0.029), compared with an observed enrollment rate of 0.369 in the analytic sample. These results suggest that feasibility-constrained ODTRs can yield improved estimated outcomes while producing recommended pathways that align with observed course-taking patterns. Consistent with the utility--feasibility trade-off discussed in the single-decision setting, more lenient propensity-score cutoffs yield slightly higher estimated values in our analysis, at the cost of recommendation feasibility. We again emphasize that, given the assumption violations discussed in Section~\ref{sec:assumptions}, these estimated values compare fitted regimes under the stated assumptions and do not measure the causal effect of implementing the recommended pathways.

\section{Design Considerations}\label{sec:design}

We discuss four practical considerations when designing the ODTRs in this section. The first is how to define feasibility. Structural constraints are claims about what is possible, so their source is institutional records rather than the analysis data, such as course catalogs, transcripts, and state prerequisite policies. For example, the rule that AP Calculus requires Pre-calculus can be written down from the course structure before any model is fitted. Normative constraints exclude options that are possible and sometimes observed, so they forfeit attainable value for the students they restrict, and the analyst should provide readers with both the argument for the rule and the authority behind it. Only relying on the statistical cutoff constraint may carry equity risk. For example, suppose students from historically underserved groups rarely take Calculus. A propensity model conditioning on their characteristics assigns low probability and therefore marks Calculus infeasible for them, so that the regime never recommends it. Several strategies might mitigate the potential equity issue. The first strategy is to report who gets trimmed based on sensitive variables like race/ethnicity, socioeconomic quartile, or school locale. Also vary the cutoff and report changes across these groups. Another remedy is to exclude protected variables (e.g., race/ethnicity, gender, SES) from the treatment model while retaining them in the outcome model if fairness is a serious concern. Doing so may reintroduce confounding bias since the fitted propensity model no longer conditions on all variables.

The second is interpretability. Flexible ML makes the recommendation mechanism hard to explain, which matters when the audience includes families and administrators. We divide interpretability into three layers: (1) key moderator discovery, (2) non-additivity (interaction) detection, (3) individual-level effect decomposition. The combination of accumulated local effect \parencite[ALE;][]{apley2020visualizing} and functional ANOVA \parencite[fANOVA;][]{hooker2004discovering,hooker2007generalized} provides a solution to address all three layers. Specifically, ALE decomposes an individual-level estimate as a grand mean, plus a first-order term per covariate, plus higher-order remainder terms. The fANOVA splits the total variance of the estimate into the share carried by each covariate, the share carried by retained interactions, and a leftover, so that it can evaluate the importance of each term based on variance share. See \textcite{hooker2007generalized} for details and \textcite{pan2026distilling} for an application. Shapley value \parencite[SHAP value;][]{lundberg2017unified} is another commonly used post hoc interpretation tool and can achieve a similar goal. 

The third concern is fairness. In addition to the equity risk carried in the feasibility constraints as discussed above, data-driven regimes can perpetuate disparities without explicitly using a protected attribute, because historical inequities, measurement bias, and structural differences across schools are encoded in the data and reproduced by any model fitting it well \parencite{suk_psychometric_2024}. \textcite{suk2024fair} incorporate student- and school-level fairness constraints into effect estimation so the resulting policies satisfy specified criteria. Evaluation should therefore extend beyond the overall value estimate to group-specific values and recommendation distributions, and to trimming rates by subgroup, where the largest disparate impact most likely lies.

Lastly, static regimes may be preferable in practice because they are straightforward to communicate and deploy in real schools, and they leave intact the peer groupings that a dynamic regime may dissolve by reassigning students across courses year by year. While causal identification in static regimes is still affected by peer interference, this is a matter of degree rather than overall validity. Because static regimes closely mimic observed cohort structures, they require far less extrapolation beyond existing data. Researchers who suspect strong positive peer effects may therefore prefer to search within the static class, or within a dynamic class constrained to preserve cohort groupings, and to report the values of the constrained and unconstrained regimes side by side. Doing so requires no additional machinery because the second component of ODTR-TMLE evaluates the learned regime by treating it as a fixed rule. The procedure in \esm{app:tech_details} applies directly to a static rule specified in advance.

\section{Limitations and Conclusions}\label{sec:limits}

A clear limitation regarding the empirical analysis is that causal assumptions are unlikely to be met in these data. Hidden variation within course categories means the treatment is a coarsening of heterogeneous exposures, interference is likely substantial, and sequential unconfoundedness is partially met at best, with unmeasured confounding most plausibly in mathematics anxiety, counselors' diagnostic information, and parental advocacy, which are not available in the data. Since the main purpose is to provide a tutorial on applying the ODTR in education, we do not conduct a sensitivity analysis for unmeasured confounding, which is the most important extension of the identification analysis. We remark that empirical analyses using causal inference should carefully examine the alignment of assumptions. On estimation, the inferential target is the value of the sample-learned rule rather than of the population-optimal regime. The learned rule is sensitive to the propensity cutoff and the \textit{a priori} specification, and ensemble methods deliver accuracy at the cost of transparency. Practitioners may consider using post hoc interpretation tools to improve interpretability. 

Three points seem worth carrying forward. First, an unconstrained ODTR model is not merely impractical here but systematically misleading, in that the largest estimated contrasts arise where the data are thinnest, so an unconstrained rule is drawn toward the least reliable region of the estimated surface, and feasibility restrictions are therefore part of a defensible estimation strategy rather than a concession to practicality. Second, the three kinds of constraint have different justifications and different consequences, and conflating them risks encoding historical patterns of access as statements about what is educationally possible. Third, a meaningful regime is not the one with the highest estimated value but one whose recommendations can be implemented, explained, and defended, and whose assumptions the analyst is prepared to state plainly. On that last point, we have tried to practice what we recommend. For example, the assumptions required for a causal reading of these results are not met in HSLS:09, and we say so rather than deferring the question to a closing paragraph. We present the empirical analyses as a demonstration that the workflow runs on real longitudinal educational data at realistic scale, and as an illustration of the design decisions any applied user will face. Researchers wanting stronger claims will need treatment definitions with less hidden variation, designs that address interference, and sensitivity analyses for unmeasured confounding. These are demanding requirements, and the honest price of moving from methodological demonstration to educational recommendation.

\section*{Electronic Supplementary Material}
\noindent ESM 1. Supplementary Sections S1--S5 (appended at the end of this preprint, starting on p.~\pageref{esm:start}).  Weighted GPA construction, variable list, technical details of ODTR-TMLE, an expert-informed course pathway, and R implementation.

% ------------------------------ REFERENCES -----------------------------------
\printbibliography

% ------------------------------ APPENDIX -------------------------------------
\appendix
\def\theHtable{A\arabic{table}}  % unique hyperref anchors (main text also has Table 1)
\section{Glossary of Technical Terms}\label{app: glossary}
\begingroup
% --- USER ADJUSTMENTS (mirrors the variable-list table settings) ---
\setlength{\LTleft}{0pt}
\setlength{\LTright}{0pt}
\setlength{\LTpost}{0pt}

{\footnotesize
\setlength{\tabcolsep}{3pt}
\renewcommand{\arraystretch}{0.9}
\linespread{0.95}\selectfont % tighten within-cell line spacing

\begin{longtable}{@{}L{3.2cm} L{12.2cm}@{}}

\caption{Glossary of key technical terms}\label{tab:glossary}\\
\toprule
\textbf{Term} & \textbf{Working definition} \\
\midrule
\endfirsthead

\caption*{\textbf{Table \thetable} \textit{(continued)}}\\
\toprule
\textbf{Term} & \textbf{Working definition} \\
\midrule
\endhead

\midrule
\multicolumn{2}{r@{}}{\textit{Continued on next page}}\\
\endfoot

\bottomrule
\multicolumn{2}{@{}p{15.4cm}@{}}{\rule{0pt}{2.2ex}\textbf{NOTE}: Working definitions paraphrase the usage in the main text and follow its notation; full technical detail appears in \esm{app:tech_details}. ESM = Electronic Supplementary Material; ODTR = optimal dynamic treatment regime; CATE = conditional average treatment effect; AIPW = augmented inverse probability weighting; LASSO = least absolute shrinkage and selection operator; ML = machine learning.}\\
\endlastfoot

Dynamic treatment regime (DTR) &
A collection of decision rules $d=(d^{(1)},\dots,d^{(K)})$, one per decision point, each mapping a student's accumulated history $h^{(k)}$ to a course in the available set. In our data, DTR is a complete four-year math course-taking plan. \textit{Dynamic} means the recommendation adapts to information realized between decisions, such as the previous year's grade. An ODTR is any regime maximizing the value. \\[2pt]

Static treatment regime &
A regime that fixes the whole course sequence from baseline information alone and does not respond to intermediate outcomes. Static regimes are easier to communicate and, when interference across students is suspected, require less extrapolation beyond observed cohort structures (Section~\ref{sec:design}). \\[2pt]

Value of a regime &
$V(d)=\mathbb E\{Y^*(d)\}$: the average outcome if the whole population followed regime $d$; the criterion by which regimes are compared. Our inferential target is $V(\hat d_{\mathrm{feas}})$, the value of the learned feasible rule treated as fixed. \\[2pt]

Backward induction (recursion) &
Solving the multi-stage problem from the last decision to the first. Because no decision follows the final one, its optimal rule is unambiguous; fixing that rule defines the continuation value that becomes the regression target for the previous stage, and the argument repeats back to Grade 9. This is how delayed effects of early courses enter early decisions. \\[2pt]

Q-function; Q-learning &
$Q^{(k)}(h^{(k)},a^{(k)})$: the expected final outcome (at the last stage) or continuation value (at earlier stages) for a student with history $h^{(k)}$ who now takes $a^{(k)}$; a nonparametric regression, with $Q$ for ``quality.'' Q-learning is the backward-recursive fitting of these regressions. \\[2pt]

Blip function &
$B^{(k)}_{j,1}(h^{(k)})=Q^{(k)}(h^{(k)},a^{(k)}_j)-Q^{(k)}(h^{(k)},a^{(k)}_1)$: the expected gain from option $a_j$ over the reference $a_1$, given optimal treatment thereafter; a stage-specific CATE. In structural nested mean models, a blip is a brief departure from the reference option at a single decision point, with the earlier record unchanged and all later decisions following the optimal rules. Rules recommend the feasible option with the largest estimated blip, with the reference having a blip of zero.  \\[2pt]

Nuisance functions &
The propensity score $e^{(k)}(a\mid h)=P(A^{(k)}=a\mid H^{(k)}=h)$ and the outcome regression $Q^{(k)}$: quantities that must be estimated to reach the target parameter but are not of substantive interest. Misspecifying them biases everything downstream, which motivates flexible estimation. \\[2pt]

Doubly robust pseudo-outcome (AIPW) &
$D_{a}(O)$: the model's prediction $\hat Q(H,a)$ of what a student would achieve under option $a$, plus an inverse-propensity-weighted correction that fires only for students actually observed under $a$. Doubly robust means consistent if either the outcome regression or the propensity model is correct. \\[2pt]

Blip pseudo-outcome &
$\tilde D_{j,1}(O)=D_{a_j}(O)-D_{a_1}(O)$: the difference between the pseudo-outcomes of option $a_j$ and the reference. Its conditional mean given the history is the blip function, so regressing blip pseudo-outcomes on $H^{(k)}$ turns a causal contrast into an ordinary regression problem. \\[2pt]

Super Learner; algorithm library &
An ensemble that combines a pre-specified library of candidate algorithms (e.g., generalized linear models, LASSO, random forest, gradient boosting) into a weighted average minimizing cross-validated error; it performs asymptotically at least as well as its best member. Penalized members such as LASSO perform automated variable selection. \\[2pt]

Cross-validation; folds &
Splitting the data into $S$ blocks (folds), training each candidate on $S-1$ folds and scoring it on the held-out fold, so that model quality is judged on data the model has not seen. \\[2pt]

Cross-fitting &
The extension of the cross-validation logic to the target parameter: nuisance functions and rules are estimated on training folds and applied only to held-out folds, so the same observations never both learn a rule and evaluate it. Without it, flexible fits carry an overfitting bias into the value estimate, and intervals are too narrow. \\[2pt]

Targeted maximum likelihood estimation (TMLE) &
A small targeted update to the initial ML regression that removes the first-order bias in the target parameter, here the value. The result is a plug-in estimator that respects parameter bounds, is doubly robust, and is asymptotically linear with a known influence function, so confidence intervals remain valid despite the use of ML. \\[2pt]

Clever covariate &
$C^{(k)}(O)$: the propensity-based term that directs the TMLE update toward the value parameter; in the longitudinal case, the indicator of having followed the learned rule through stage $k$ divided by the cumulative propensities of the options taken. \\[2pt]

Efficient influence function (EIF) &
$\phi(O)$: the function whose empirical variance gives the estimator's asymptotic variance; the basis of the standard errors and Wald-type 95\% confidence intervals reported for the value of the learned regime. \\[2pt]

Feasible set; feasibility constraints &
$\mathcal A^{(k)}_{\mathrm{feas}}(h^{(k)})\subseteq\mathcal A^{(k)}$: the options a rule may recommend given history $h^{(k)}$, combining \textit{a priori} structural and normative rules with a propensity-score threshold that removes empirically unsupported options. Restricted positivity (A4) is required only over feasible options, and $\mathcal D_{\mathrm{feas}}$ (Equation~\eqref{eq:feasible_class}) is the set of regimes whose rule at every stage recommends only options in the feasible set of the history at hand, a set of regimes rather than of courses. \\

\end{longtable}
}
\endgroup

% =============================================================================
% ELECTRONIC SUPPLEMENTARY MATERIAL 1
% Published by the journal as a separate file (ESM1.pdf); appended here so the
% preprint is self-contained. Title block as in ESM.tex.
% =============================================================================
% apa6's \appendix replaces \section with an unnumbered "Appendix" heading that
% ignores the section counter and \section*; restore the regular heading so the
% ESM sections are numbered S1--S5 and its reference list gets a normal heading.
\makeatletter
\let\section\old@apaSix@section
\makeatother
% unique hyperref anchors for the S-numbered sections, equations and tables
\def\theHsection{S\arabic{section}}
\def\theHequation{S\arabic{equation}}
\def\theHtable{S\arabic{table}}
\def\theHfigure{S\arabic{figure}}

\clearpage
\phantomsection\label{esm:start}
\pdfbookmark[0]{Electronic Supplementary Material 1}{esm1}
\begin{center}
\textbf{Electronic Supplementary Material 1 (ESM 1)}\par
\textit{Estimating Optimal Dynamic Treatment Regimes for Personalized Education: A Tutorial and Applications with Machine Learning}\par
Chenguang Pan, Yuxuan Li, Thomas Harris, and Youmi Suk
\end{center}

% separate reference list for the ESM, as in the stand-alone ESM1.pdf
\begin{refsection}
% let long \texttt identifiers (e.g. tmle3_mopttx_blip_revere) break after "_"
% so the justified ESM paragraphs do not get huge word gaps
\let\ESMorigus\_
\renewcommand{\_}{\ESMorigus\allowbreak}
\input{suppl}   % Sections S1--S5

\printbibliography
\end{refsection}

\end{document}

%% file: suppl.tex
% --- START OF suppl.tex ---
% Note: No \documentclass, \usepackage, or \begin{document} here.

% Reset counters for the appendix
\setcounter{secnumdepth}{1}
\setcounter{equation}{0}
\setcounter{figure}{0}
\setcounter{table}{0}
\setcounter{section}{0}

% Rename counters to S1, S2, etc.
\renewcommand{\theequation}{S\arabic{equation}}
\renewcommand{\thefigure}{S\arabic{figure}}
\renewcommand{\thesection}{S\arabic{section}}
\renewcommand{\thetable}{S\arabic{table}}

\section{Construction of Weighted GPA} \label{app:GPA}
The raw transcript GPA cannot represent the student's math progress. For example, a letter grade A in Pre-calculus may indicate higher math capability than an A in Algebra I. Therefore, we assign each course level a rigor weight $w$: 0.5 for Basic and Other Mathematics, 0.8 for Pre-Algebra, 1.0 for Algebra I and Geometry, 1.15 for Algebra II, Advanced Mathematics, and Probability and Statistics, 1.3 for AP/IB Mathematics and Pre-calculus, 1.5 for Calculus, and 1.6 for AP Calculus. Writing $g_c$ for the grade points earned in course $c$, $k_c$ for credits earned and $k_c^{\mathrm{pot}}$ for credits attempted,
\begin{equation}\label{eq:wgpa}
\mathrm{wGPA} = \frac{\sum_c w_c\, g_c\, k_c}{\sum_c k_c^{\mathrm{pot}}},
\end{equation}
following the NCES transcript convention for the denominator, which does not contain the weights. Two consequences follow and should be stated plainly. First, wGPA runs from 0 to 6.4 rather than 0 to 4, and it is a joint index of grade performance and curricular rigor. Second, wGPA increases with course level at fixed grades, so maximizing it rewards acceleration at an explicit exchange rate given by the weight ratio. For example, Algebra II is preferred to Algebra I whenever the expected grade falls by less than 13\%, and AP Calculus is preferred to Algebra I whenever it falls by less than 37.5\%, so that a C in AP Calculus (3.20) outranks a B in Algebra I (3.00). 

The measure is cumulative only at the final decision point. The record entering the history at Grades 10, 11, and 12 is that grade's own weighted mathematics GPA, computed over that year's courses alone, whereas the final outcome aggregates all four years. We use the cumulative version as the final outcome so that the optimization target reflects performance over the entire course-taking sequence rather than the final year alone. A Grade 12--only outcome would credit early decisions only through their effect on a single year's grades.

\newpage
\section{Variable List} \label{app:var_list}

We provide the full variable list in Table~\ref{app:tab:vars}, which reports each variable’s description, analytic role, and data source or construction rule. Variable names are defined for ease of implementation and communication. We use Y1 as the outcome in the single-decision OTR model, Y4 as the outcome in the multi-stage math achievement model, and Y6 as the outcome in the multi-stage college enrollment model. The treatment variables A1 through A4 denote the highest math course level completed in grades 9 through 12, respectively. Table~1 lists the categories for each treatment.

% --- USER ADJUSTMENTS ---
\setlength{\LTleft}{0pt}
\setlength{\LTright}{0pt}
\setlength{\LTpost}{0pt}

{\footnotesize
\setlength{\tabcolsep}{1pt}
\renewcommand{\arraystretch}{0.7}
\linespread{0.95}\selectfont % tighten within-cell line spacing

\begin{longtable}{@{}L{1.6cm} L{5.8cm} C{1.0cm} C{1.0cm} C{1.0cm} C{1.0cm} L{4.0cm}@{}}

\caption{Variables used in this study from HSLS:09 data}\label{app:tab:vars}\\
\toprule
\textbf{Variable} & \textbf{Description} & \textbf{Gr9} & \textbf{Gr10} & \textbf{Gr11} & \textbf{Gr12} & \textbf{Source} \\
\midrule
\endfirsthead

\toprule
\textbf{Variable} & \textbf{Description} & \textbf{Gr9} & \textbf{Gr10} & \textbf{Gr11} & \textbf{Gr12} & \textbf{Source} \\
\midrule
\endhead

\midrule \multicolumn{7}{r}{\textit{Continued on next page}}\\
\endfoot

\bottomrule
\endlastfoot

  Y1 & Weighted math GPA (Grade 9) & & C;M & C;M & C;M & Constructed from transcripts dataset \\
  \hdashline[0.5pt/2pt]
  Y2 & Weighted math GPA (Grade 10) & & & C;M & C;M & Constructed from transcripts dataset \\
  \hdashline[0.5pt/2pt]
  Y3 & Weighted math GPA (Grade 11) & & & & C;M & Constructed from transcripts dataset \\
  \hdashline[0.5pt/2pt]
  Y4 & Cumulative weighted math GPA (Grade 12) & & & & & Constructed from transcripts dataset \\
  \hdashline[0.5pt/2pt]
  Y6 & Bachelor's degree program enrollment indicator & & & & & X3PROGLEVEL \\
  \hdashline[0.5pt/2pt]
  A1 & Highest-level math course completed in Grade 9 & & C & C & C & Constructed from transcripts dataset \\
  \hdashline[0.5pt/2pt]
  A2 & Highest-level math course completed in Grade 10 & & & C & C & Constructed from transcripts dataset \\
  \hdashline[0.5pt/2pt]
  A3 & Highest-level math course completed in Grade 11 & & & & C & Constructed from transcripts dataset \\
  \hdashline[0.5pt/2pt]
  A4 & Highest-level math course completed in Grade 12 & & & & & Constructed from transcripts dataset \\
  \hdashline[0.5pt/2pt]
  L4 & Weighted math score (pre-high school self-reported GPA) & C;M & C;M & C;M & C;M & S1M8GRADE \\
  \hdashline[0.5pt/2pt]
  S1 & Student race/ethnicity & C & C & C & C & X1RACE \\
  \hdashline[0.5pt/2pt]
  S2 & Proportion of Black students in school & C & C & C & C & X1SCHBLACK \\
  \hdashline[0.5pt/2pt]
  X001* & Parents' education level & C & C & C & C & X1PAREDU | X2PAREDU \\
  \hdashline[0.5pt/2pt]
  X002 & Student gender & C & C & C & C & X1SEX \\
  \hdashline[0.5pt/2pt]
  X003* & Perceived math utility & C & C & C & C & X1MTHUTI | X2MTHUTI \\
  \hdashline[0.5pt/2pt]
  X004* & School climate index (higher values indicate better climate) & & C;M & C;M & C;M & X1SCHOOLCLI | X2SCHOOLCLI \\
  \hdashline[0.5pt/2pt]
  X005* & Math identity & C;M & C;M & C;M & C;M & X1MTHID | X2MTHID \\
  \hdashline[0.5pt/2pt]
  X006 & Student age & C & C & C & C & X1STDOB \\
  \hdashline[0.5pt/2pt]
  X007* & Math self-efficacy & C;M & C & C & C & X1MTHEFF | X2MTHEFF \\
  \hdashline[0.5pt/2pt]
  X008* & Interest in math course & & C;M & C;M & C;M & X1MTHINT | X2MTHINT\_R \\
  \hdashline[0.5pt/2pt]
  X009 & Sense of school belonging & & C & C;M & C & X1SCHOOLBEL \\
  \hdashline[0.5pt/2pt]
  X010 & Participation in math activities & C & C & C & C & S1MCAMP + S1MCLUB + S1MCOMPETE + S1MTUTOR \\
  \hdashline[0.5pt/2pt]
  X011* & Parents' educational expectations for student & C & C & C & C & X1PAREDEXPCT | X2PAREDEXPCT \\
  \hdashline[0.5pt/2pt]
  X012* & Student's educational expectations & C & C & C & C & X1STUEDEXPCT | X2STUEDEXPCT \\
  \hdashline[0.5pt/2pt]
  X013* & Parents' STEM occupation status (1 = at least one parent in STEM) & C & C & C & C & X1PAR1OCC\_STEM1 + X1PAR2OCC\_STEM1 | X2PAR1OCC\_STEM1 + X2PAR2OCC\_STEM1 \\
  \hdashline[0.5pt/2pt]
  X014* & School hosts math/science fairs or competitions & & C & C & C & A1MTHSCIFAIR | A2MTHSCIFAIR \\
  \hdashline[0.5pt/2pt]
  X016\_R & Student-to-math-teacher ratio & C;M & C;M & C;M & C;M & A2HSSIZE + A1TOTMTCHRS \\
  \hdashline[0.5pt/2pt]
  X017 & AP course offering availability & & C & C & C & A1OFFERAP \\
  \hdashline[0.5pt/2pt]
  X018 & Math course graduation requirements & C & C & C & C & A1MTHREQS \\
  \hdashline[0.5pt/2pt]
  X019 & School geographic region & C & C & C & C & X1REGION \\
  \hdashline[0.5pt/2pt]
  X021 & Summer transition program availability & C & C & C & C & A1G9SUMMER \\
  \hdashline[0.5pt/2pt]
  X022 & Counselor assistance in transition & C & C & C & C & C1TRANSCRS + C1TRANSCNSL + C1TRANPLCY + C1TRANPRNT + C1TRANPRES + C1TRANCOTH \\
  \hdashline[0.5pt/2pt]
  X027 & School assistance in transition & C & C & C & C & C1TRANSTUDPR, C1TRANSTFFPR, C1TRANVISIT, C1TRANCLASS, C1TRANADMIN, C1TRANTCHRS, C1TRANBUDDY, C1TRANLRNCOM, C1TRANSUMMER, C1TRANFALL, C1TRANSOTH \\
  \hdashline[0.5pt/2pt]
  X033 & Program for underrepresented students in math/science & & C & C & C & C1PURSUE \\
  \hdashline[0.5pt/2pt]
  X034 & Math competency test requirement & & C & C & C & C1MCOMPTST \\
  \hdashline[0.5pt/2pt]
  X035 & Universal 9th-grade math placement & C & C & C & C & C1G9MSAME \\
  \hdashline[0.5pt/2pt]
  X036 & Importance of placement tests for 9th-grade assignment & C & C & C & C & C1G9MPLACTST \\
  \hdashline[0.5pt/2pt]
  X037 & Universal math placement (post-9th grade) & & C & C & C & C1UPPERMSAME \\
  \hdashline[0.5pt/2pt]
  X038 & School locale (city, town, rural, suburb) & C & C & C & C & X1LOCALE \\
  \hdashline[0.5pt/2pt]
  X039 & School sector (public, private, Catholic) & C & C & C & C & X1CONTROL \\
  \hdashline[0.5pt/2pt]
  X040 & Remedial math course availability & C & C & C & C & X1REMMATH \\
  \hdashline[0.5pt/2pt]
  X041 & Alignment of school and state math requirements & C & C & C & C & A1MTHSTREQ \\
  \hdashline[0.5pt/2pt]
  X042* & Student household SES & C & C & C;M & C & X1SES | X2SES \\
  \hdashline[0.5pt/2pt]
  X043* & Standardized algebra assessment score & & C;M & C;M & C;M & X1TXMTSCOR | X2TXMTSCOR \\
  \hdashline[0.5pt/2pt]
  X044 & School engagement & & C;M & C;M & C & X1SCHOOLENG \\
  \hdashline[0.5pt/2pt]
  X045 & School academic support services & & C & C & C & C1TUTOR, C1STAFF, C1PULLOUT, C1CREDREC, C1HOMEWORK, C1OUTSIDE, C1OTHRASSIST \\
  \hdashline[0.5pt/2pt]
  X046 & External motivation for math & & C & C & C & S1MREASREQ + S1MREASPAR + S1MREASTCHR + S1MREASCNSL \\
  \hdashline[0.5pt/2pt]
  X047 & Understanding of math assignments & & C & C & C & S1MUNDERST \\
  \hdashline[0.5pt/2pt]
  X048 & School change status (previous academic year) & & C & C & C & S1SCH0809 \\
  \hdashline[0.5pt/2pt]
  X050 & Math teacher's attitude toward student & & C & C & C & S1MTCHRESPCT, S1MTCHVALUES, S1MTCHFAIR, S1MTCHCONF, S1MTCHMISKE \\
  \hdashline[0.5pt/2pt]
  X051* & Math teacher's pedagogical skills & & & C & C & S1MTCHINTRST, S1MTCHEASY | S2MTCHINTRST, S2MTCHEASY \\
  \hdashline[0.5pt/2pt]
  X052 & Counselor quality & & C & C & C & C1CLEARNING, C1CBELIEVE, C1CWORKHARD, C1CGIVEUP, C1CCARE, C1CEXPECT \\
  \hdashline[0.5pt/2pt]
  X053 & School motivation & & & & C & X2BEHAVEIN \\
  \hdashline[0.5pt/2pt]
  X054 & Math class effort scale & & & & C;M & X2MEFFORT \\
  \hdashline[0.5pt/2pt]
  X055 & High school problems scale & & & & C & X2PROBLEM \\
  \hdashline[0.5pt/2pt]
  X060* & Science identity & & C;M & C;M & C;M & X1SCIID \\

\end{longtable}

% Add notes after the table
\scriptsize
\small
\noindent
\textbf{NOTE}: Variable names are chosen for ease of implementation and communication. The outcomes Y1--Y4 and Y6, and treatments A1--A4 are constructed from the restricted-use transcript file \texttt{stu\_course.sas7bdat} in HSLS:09. Variables marked with an asterisk are time-varying and are measured in grades 9 and 11. For each variable, we indicate its analytic role at each grade from 9 to 12 (Gr9--Gr12), where "C" denotes a confounder and "M" denotes a moderator. Some source variables are available only in the restricted-use data. \newline
\textbf{SOURCE}: Department of Education, National Center for Education Statistics, High School Longitudinal Study of 2009 (HSLS:09).

}

\newpage
\section{Explaining ODTR-TMLE Steps using Mathematical expressions}\label{app:tech_details}

We utilize the TMLE-based method to estimate the ODTRs and the value of the learned rules \parencite[]{van2015target}. The procedure has two phases. First, we learn optimal regimes by backward induction with stage-specific regressions as outlined in Section~4 of the main text. Second, we estimate and update the value of the learned regimes with cross-validated TMLE \parencite{zheng2011cvtmle}. Comprehensive R scripts to implement this workflow are publicly available at \url{https://github.com/cgpan/feasibleODTR}.

Before describing the backward-induction algorithm, we clarify how feasibility constraints are incorporated into regime learning and value estimation. We therefore restrict regime learning and evaluation to feasible options. We consider two constructions of $\mathcal A^{(k)}_{\mathrm{feas}}(h^{(k)})$. First, \textit{a priori} constraints encode structural and normative constraints on option availability for subpopulations. Second, propensity-score constraints exclude options with small estimated probability \parencite{van2007realistic}:
\[
\mathcal A^{(k)}_{\mathrm{feas}}(h^{(k)})=
\left\{a\in\mathcal A^{(k)}:\hat e^{(k)}(a\mid h^{(k)})\ge \alpha^{(k)}\right\},
\]
where $\alpha^{(k)}$ denotes a prespecified threshold. Under this restriction, assumption (A4) requires positivity only for options in $\mathcal A^{(k)}_{\mathrm{feas}}(h^{(k)})$ \parencite{van2007realistic}.

\subsection{Phase 1: Feasible ODTR Learning}
Given these feasible option sets, we next describe how the ODTR is learned by backward induction. For decision point $k\in\{1,\dots,K\}$, let $\mathcal A^{(k)}=\{a^{(k)}_1,\dots,a^{(k)}_{J_k}\}$ denote the $J_k$ treatment options available at this decision point. Let
\[
e^{(k)}(a\mid h)=P(A^{(k)}=a\mid H^{(k)}=h)
\]
denote the stage-$k$ treatment model. Let $Y^{(K)}=Y$, and for $k<K$, let $Y^{(k)}$ denote the pseudo-outcome summarizing the optimal value from later decision points, defined recursively below. The stage-$k$ outcome regression is
\[
Q^{(k)}(h,a)=\mathbb E\!\left(Y^{(k)} \mid H^{(k)}=h, A^{(k)}=a\right).
\]

At the final decision point $K$, we estimate $\hat Q^{(K)}(h^{(K)},a^{(K)})$ and $\hat e^{(K)}(a^{(K)}\mid h^{(K)})$ with Super Learner. For each option $a\in\mathcal A^{(K)}$, we form an augmented inverse probability weighted outcome \parencite{Robins1994Estimation}:
\begin{equation}\label{eq:aipw_K}
D^{(K)}_{a}(O)
=
\frac{\mathbbm{1}\{A^{(K)}=a\}}{\hat e^{(K)}(a\mid H^{(K)})}
\left\{Y-\hat Q^{(K)}(H^{(K)},a)\right\}
+\hat Q^{(K)}(H^{(K)},a).
\end{equation}
We select a reference option $a^{(K)}_1$ and define the blip pseudo-outcome for option $a^{(K)}_j$ relative to $a^{(K)}_1$ as:
\begin{equation}\label{eq:blip_K}
\tilde D^{(K)}_{j,1}(O)=D^{(K)}_{a^{(K)}_j}(O)-D^{(K)}_{a^{(K)}_1}(O),\qquad j=2,\dots,J_K .
\end{equation}
Then, the corresponding blip function $B^{(K)}_{j,1}(h^{(K)})=\mathbb E\{\tilde D^{(K)}_{j,1}(O)\mid H^{(K)}=h^{(K)}\}$ is estimated by regressing $\tilde D^{(K)}_{j,1}$ on $H^{(K)}$ \parencite{robins2004optimal}. Let $\hat B^{(K)}_{j,1}(h^{(K)})$ denote the fitted function. We set $\hat B^{(K)}_{j,1}(h^{(K)}) = -\infty$ for any option $a^{(K)}_j \notin \mathcal A_{\mathrm{feas}}^{(K)}(h^{(K)})$ so that we do not recommend these infeasible options. The learned feasible rule at stage $K$ then takes the form
\begin{equation}\label{eq:rule_K}
\hat d_{\mathrm{feas}}^{(K)}(h^{(K)})=\mathop{\mathrm{argmax}}_{a^{(K)}_j \in \mathcal A_{\mathrm{feas}}^{(K)}(h^{(K)})}\hat B^{(K)}_{j,1}(h^{(K)}),\qquad \hat B^{(K)}_{1,1}(h^{(K)})\equiv 0.
\end{equation}
% \begin{equation}\label{eq:rule_K}
% \hat d_{\mathrm{feas}}^{(K)}(h^{(K)})
% =
% \begin{cases}
% a^{(K)}_1, &
% \max\limits_{a^{(K)}_j \in \mathcal A_{\mathrm{feas}}^{(K)}(h^{(K)}),\, j \ge 2}
% \hat B^{(K)}_{j,1}(h^{(K)}) \le 0, \\[6pt]
% \mathop{\mathrm{argmax}}\limits_{a^{(K)}_j \in \mathcal A_{\mathrm{feas}}^{(K)}(h^{(K)}),\, j \ge 2}
% \hat B^{(K)}_{j,1}(h^{(K)}), &
% \text{otherwise}.
% \end{cases}
% \end{equation}
We also define an estimated downstream value function at stage $K$ as
\[
\hat V^{(K)}(h^{(K)})=\hat Q^{(K)}\!\left(h^{(K)},\hat d^{(K)}_{\mathrm{feas}}(h^{(K)})\right).
\]

For earlier stages $k=K-1,\dots,1$, we similarly construct a pseudo-outcome that carries information from later optimal decisions. We first set $Y^{(k)}=\hat V^{(k+1)}(H^{(k+1)})$ and estimate $\hat Q^{(k)}(h^{(k)},a^{(k)})$ and $\hat e^{(k)}(a^{(k)}\mid h^{(k)})$. Next, we form $D^{(k)}_{a}(O)$ by replacing $Y$ with $Y^{(k)}$ in Equation~\eqref{eq:aipw_K} and regress the resulting blip pseudo-outcomes on $H^{(k)}$. We then define $\hat d^{(k)}(h^{(k)})$ by the feasible version of Equation~\eqref{eq:rule_K}. This backward recursion yields the learned feasible ODTR $\hat d=(\hat d^{(1)}_{\mathrm{feas}},\dots,\hat d^{(K)}_{\mathrm{feas}})$. Cross-fitting is applied for all stages to limit overfitting bias when the same data inform both rule learning and value estimation. Each fold uses training data for nuisance estimation and rule learning, and it uses held-out data for prediction and value targeting.

\subsection{Phase 2: Value estimation using TMLE}
After learning the feasible regime, we estimate the value $V(\hat d_{\mathrm{feas}})=\mathbb E\{Y^*(\hat{d}_{\mathrm{feas}})\}$ using cross-validated TMLE. For each fold, we treat the learned regime $\hat d$ and nuisance fits from the training split as fixed functions on the validation split. Let $Y^{(K)}=Y$ and let $Y^{(k)}$ for $k<K$ denote the fold-specific pseudo-outcome that equals the updated value estimate from stage $k+1$. Let $\hat Q^{(k)}_{\mathrm{init}}(h,a)$ denote the initial estimate of $\mathbb E(Y^{(k)}\mid H^{(k)}=h,A^{(k)}=a)$ on the validation split. When $Y^{(k)}$ is continuous, we first place the outcome and the initial regression on the unit interval before applying the logistic fluctuation step. This normalization is a numerical device that puts both quantities on the scale required by the logistic update. We skip this step for binary outcomes. Specifically,
\begin{align}\label{eq:minmax}
Y^{(k)}_{\mathrm{sc}} = \frac{Y^{(k)} - Y^{(k)}_{\min}}{Y^{(k)}_{\max}-Y^{(k)}_{\min}},
\qquad
\hat{Q}^{(k)}_{\mathrm{init,sc}}(h,a) = \frac{\hat{Q}^{(k)}_{\mathrm{init}}(h,a) - Y^{(k)}_{\min}}{Y^{(k)}_{\max}-Y^{(k)}_{\min}},
\end{align}
where $Y^{(k)}_{\min}$ and $Y^{(k)}_{\max}$ denote the stage-specific lower and upper bounds used for scaling, chosen so that the observed $Y^{(k)}$ falls within them. The same bounds are applied to both quantities. In implementation, the scaled initial predictions are additionally truncated to $[\varepsilon,\,1-\varepsilon]$ with $\varepsilon = 0.005$ so that the logit in Equation~\eqref{eq:fluc_model} remains finite. The scaled outcome itself is not truncated.

We then construct the clever covariate for stage $k$ as
\begin{equation}\label{eq:clever_cv}
C^{(k)}(O)=\prod_{t=1}^{k}
\frac{\mathbbm{1}\!\left\{A^{(t)}=\hat d^{(t)}_{\mathrm{feas}}(H^{(t)})\right\}}
{\hat e^{(t)}\!\left(A^{(t)}\mid H^{(t)}\right)}.
\end{equation}
Using this clever covariate, we fit a logistic fluctuation model on the validation split with offset $\text{logit}\{\hat Q^{(k)}_{\mathrm{init,sc}}(H^{(k)},A^{(k)})\}$:
\begin{equation}\label{eq:fluc_model}
\text{logit}\{\mathbb E(Y^{(k)}_{\mathrm{sc}}\mid H^{(k)},A^{(k)})\}
=
\text{logit}\{\hat Q^{(k)}_{\mathrm{init,sc}}(H^{(k)},A^{(k)})\}
+\epsilon^{(k)} C^{(k)}(O).
\end{equation}
Let $\hat\epsilon^{(k)}$ denote the fitted coefficient. The scaled outcome can be updated as
\begin{equation}\label{eq:update}
\hat Q^{(k)}_{\mathrm{upd,sc}}(H^{(k)},A^{(k)})
=
\text{expit}\!\left(
\text{logit}\{\hat Q^{(k)}_{\mathrm{init,sc}}(H^{(k)},A^{(k)})\}
+\hat\epsilon^{(k)} C^{(k)}(O)
\right),
\end{equation}
and we then rescale to obtain $\hat Q^{(k)}_{\mathrm{upd}}$ on the original outcome scale.

The updated regression is used to define the value carried backward to the previous stage:
\begin{equation}\label{eq: pseudo_outcome}
\hat V^{(k)}(h^{(k)})=\hat Q^{(k)}_{\mathrm{upd}}\!\left(h^{(k)},\hat d^{(k)}_{\mathrm{feas}}(h^{(k)})\right).
\end{equation}
We then set $Y^{(k-1)}=\hat V^{(k)}(H^{(k)})$ for the next step, and this recursion proceeds from $k=K$ down to $k=1$.

At the end of the backward recursion, $\hat V^{(1)}(X^{(1)})$ represents the estimated conditional expected final outcome under the entire learned feasible regime given the baseline covariates. This is because each pseudo-outcome $\hat V^{(k)}$ carries forward the estimated optimal value from later decision points, so the stage-$1$ value function already incorporates all subsequent treatment decisions and outcomes. By the law of iterated expectations, the regime value is therefore obtained by averaging this stage-$1$ conditional value over the baseline distribution. Accordingly, the fold-specific value estimate is the empirical mean of $\hat V^{(1)}(X^{(1)})$ on the validation split, and the cross-validated value estimator averages these estimates across folds:
\begin{equation}\label{eq:value}
\hat V(\hat{d}_\mathrm{feas})=\frac{1}{n}\sum_{i=1}^n \hat V^{(1)}(X^{(1)}_i).
\end{equation}

Finally, we conduct inference for the estimated value of the learned feasible regime through the corresponding EIF. For the fixed learned regime on each validation split, the EIF for the value estimator has the form:
\begin{equation}\label{eq:EIFs}
\phi(O)
=
\sum_{k=1}^{K} C^{(k)}(O)\Bigl\{Y^{(k)}-\hat Q^{(k)}_{\mathrm{upd}}(H^{(k)},A^{(k)})\Bigr\}
+\hat V^{(1)}(X^{(1)})-\hat V(\hat{d}_{\mathrm{feas}}),
\end{equation}
with $Y^{(K)}=Y$ \parencite{vanderlaan2006tml}. We estimate the asymptotic variance by $\hat\sigma^2=n^{-1}\sum_{i=1}^n \phi(O_i)^2$ and report a Wald-type $95\%$ interval
\begin{equation}\label{eq: inference}
   \hat V(\hat d)\pm 1.96\,\frac{\hat\sigma}{\sqrt{n}}. 
\end{equation}
When $K=1$, this procedure reduces to the single-decision OTR-TMLE workflow.

Practitioners may wonder how the backward induction differs from simply predicting future achievement. A regression of the final outcome on the current course ranks options by what students who took them subsequently did, so an option inherits the follow-on choices of its historical takers. The recursion instead evaluates each option against the best feasible continuation. For example, standard regression might favor Probability and Statistics over Pre-calculus because past takers of Probability and Statistics achieved a higher average final cumulative GPA. However, backward induction can still prefer Pre-calculus because it unlocks AP Calculus in Grade 12, and the best continuation through Pre-calculus can carry a higher expected cumulative GPA than any route available after Probability and Statistics.

\newpage
\section{A Theory-driven, Expert-informed Math Course Pathway}\label{app:theory-driven}

In addition to the data-driven ODTR model, we provide a theory-driven, expert-informed math course pathway framework that maps a small set of student characteristics to a four-year course sequence. We first develop a brief pathway survey to collect the students’ educational goals (mathematics achievement vs college enrollment), current ninth-grade math course, perception of math identity (MI), anticipated MI change, perception of STEM professional identity, and willingness to accelerate via two math courses in one academic year. We then develop math course recommendation models for each educational goal by taking the collected information as input.

The first model, the Mathematics Achievement (MA) model, emphasizes both math performance (as indicated by a weighted math GPA) and math identity development. In contrast, the College Enrollment (CE) model focuses on the likelihood of college enrollment and places greater weight on math elective courses. We use the Math Identity Index \parencite[MII;][]{bohrnstedt2020mathematics} to indicate the MI by summing two survey items into a 0–6 scale, and then we assign students to low, mid, and high MII groups. We use a Professional Identity Index \parencite[PII;][]{mcdonald2019single} that captures STEM professional identity alignment and stratifies it into low, mid, and high categories. The models also offer an acceleration option that recommends two math courses in the same academic year, if feasible.

The pathway logic respects course prerequisites and developmental timing. Both models treat ninth-grade placement as an input rather than a decision, and they anchor recommendations in four fundamental course categories (Basic Math/Pre-algebra, Algebra I, Geometry, Algebra II) as entry points. We also encode the regular math course requirements and equivalences into two models, such as Algebra I as a prerequisite for Geometry and Geometry as a prerequisite for Algebra II and Probability and Statistics. Finally, the full pathway system appears as transparent decision rules that map MII subgroup, PII level, and acceleration status to grade-specific course choices, which supports both interpretability and direct comparison with data-driven ODTR recommendations. Details and demonstrations are available at \url{https://youmilab.ai/personalized-plan}.

The framework is developed by the author team, which includes a mathematics education specialist and draws on identity-based motivation theory and standard course prerequisite structures. It is expert-informed in this sense rather than elicited from an external expert panel.

\newpage

\newpage
\section{Applying ODTR-TMLE in R}\label{app:R_app}

We conduct the data analysis in R \parencite[version 4.3.0;][]{R2025} using the \texttt{tmle3} \parencite[][]{jeremy2021tmle3}, \texttt{tmle3mopttx} \parencite[][]{malenica2022tmle3mopttx}, and \texttt{sl3} \parencite[][]{Jeremy2023sl3} packages. These packages provide a convenient workflow for single-decision OTR estimation, but they offer limited flexibility for user-defined modifications, such as applying a custom propensity-score cutoff to remove infeasible treatment options or imposing external feasibility constraints on recommendations. To address this limitation, we develop two functions, \texttt{extract\_tmle\_components} and \texttt{Optimizer}, to extract the key intermediate quantities and perform value updating under the feasible rule in one sitting.

In the single-stage OTR task, after completing Steps 1 and 2 with the \texttt{tmle3\_mopttx\_blip\_revere} function from \texttt{tmle3mopttx} package, \texttt{extract\_tmle\_components} retrieves the matrices of counterfactual outcomes, propensity scores, and blips needed to implement customized feasibility constraints. Given these quantities and the learned feasible rule, \texttt{Optimizer} takes the observed treatment assignments, observed outcomes, estimated counterfactual outcomes, propensity scores, and feasible recommendations as inputs, and returns the updated value estimate, efficient influence functions (EIFs), and standard error (SE). 

Existing R software does not directly implement the full workflow for multi-stage ODTR learning with feasibility constraints and subsequent value updating under the learned ODTRs. For example, the \texttt{ltmle} package supports longitudinal TMLE for user-specified dynamic interventions, but it is not designed to learn ODTRs and then evaluate the value under the learned regime. Therefore, we build a customized ODTR workflow using the same \texttt{tlverse} components as in the single-decision section, supplemented with additional stage-specific operations for backward induction and value updating. We provide a comprehensive demo code based on simulated data in the GitHub repository.

%% file: ref.bib
@techreport{burkam2003mathematics,
  title={Mathematics, foreign Language, and science coursetaking and the NELS: 88 transcript data},
  author={Burkam, David T and Lee, Valerie E},
  year={2003},
  institution={National Center for Education Statistics},
  type={Working Paper},
  number={2003-01},
  url={https://eric.ed.gov/?id=ED566861}
}

@misc{pan2026distilling,
	title = {Distilling {Black}-{Box} {Machine} {Learning} into a {Small}, {Self}-{Explaining} {Language} {Model} for {Learning} {Analytics}},
	url = {http://arxiv.org/abs/2608.21165},
	doi = {10.48550/arXiv.2608.21165},
	urldate = {2026-09-22},
	publisher = {arXiv},
	author = {Pan, Chenguang and Meng, Airui and Suk, Youmi},
	month = aug,
	year = {2026},
	note = {arXiv:2608.21165 [cs.HC]}
}

@article{chernozhukov2018double,
  title={Double/debiased machine learning for treatment and structural parameters},
  author={Chernozhukov, Victor and Chetverikov, Denis and Demirer, Mert and Duflo, Esther and Hansen, Christian and Newey, Whitney and Robins, James},
  journal={The Econometrics Journal},
  volume={21},
  number={1},
  pages={C1--C68},
  year={2018},
  doi={10.1111/ectj.12097}
}

@techreport{dalton2007advanced,
  title={Advanced mathematics and science coursetaking in the spring high school senior classes of 1982, 1992, and 2004},
  author={Dalton, Ben and Ingels, Steven J and Downing, J and Bozick, R},
  year={2007},
  institution={National Center for Education Statistics},
  type={Statistical analysis report (Working Paper No. 2007-312)},
  url={https://eric.ed.gov/?id=ED497754}
}

@article{dumont2023promise,
  title={On the promise of personalized learning for educational equity},
  author={Dumont, Hanna and Ready, Douglas D},
  journal={npj Science of Learning},
  volume={8},
  number={1},
  pages={26},
  year={2023},
  doi={10.1038/s41539-023-00174-x}
}

@techreport{ingels2011hsls,
  title={High School Longitudinal Study of 2009 (HSLS: 09): Base-Year Data File Documentation},
  author={Ingels, Steven J and Pratt, D J and Herget, D R and Burns, L J and Dever, J A and Ottem, R and others},
  year={2011},
  institution={National Center for Education Statistics},
  number={NCES 2011-328},
  url={https://eric.ed.gov/?id=ED566098}
}

@article{murphy2007developing,
  title={Developing adaptive treatment strategies in substance abuse research},
  author={Murphy, Susan A and Lynch, Kevin G and Oslin, David and McKay, James R and TenHave, Thomas},
  journal={Drug and Alcohol Dependence},
  volume={88},
  pages={S24--S30},
  year={2007},
  doi={10.1016/j.drugalcdep.2006.09.008}
}

@book{tsiatis2019dynamic,
  title={Dynamic treatment regimes: Statistical methods for precision medicine},
  author={Tsiatis, Anastasios A and Davidian, Marie and Holloway, Shannon T and Laber, Eric B},
  year={2019},
  edition={1st},
  publisher={Chapman and Hall/CRC},
  doi={10.1201/9780429192692}
}

@article{suk2023designing,
  title={Designing optimal, data-driven policies from multisite randomized trials},
  author={Suk, Youmi and Park, Chan},
  journal={Psychometrika},
  volume={88},
  pages={1171--1196},
  year={2023},
  doi={10.1007/s11336-023-09937-2}
}

@article{vanderlaan2007super,
  title={Super learner},
  author={{van der Laan}, Mark J and Polley, Eric C and Hubbard, Alan E},
  journal={Statistical Applications in Genetics and Molecular Biology},
  volume={6},
  number={1},
  year={2007},
  doi={10.2202/1544-6115.1309}
}

@article{van2007realistic,
	title = {Causal {Effect} {Models} for {Realistic} {Individualized} {Treatment} and {Intention} to {Treat} {Rules}},
	volume = {3},
	issn = {2194-573X},
	url = {https://pmc.ncbi.nlm.nih.gov/articles/PMC2613338/},
	doi = {10.2202/1557-4679.1022},
	number = {1},
	urldate = {2026-02-21},
	journal = {The international journal of biostatistics},
	author = {{van der Laan}, Mark J and Petersen, Maya L},
	year = {2007},
	pages = {Article3}
}

@article{zhang2012estimating,
	title = {Estimating optimal treatment regimes from a classification perspective},
	volume = {1},
	copyright = {Copyright © 2012 John Wiley \& Sons, Ltd.},
	issn = {2049-1573},
	url = {https://onlinelibrary.wiley.com/doi/abs/10.1002/sta.411},
	doi = {10.1002/sta.411},
	language = {en},
	number = {1},
	urldate = {2026-02-21},
	journal = {Stat},
	author = {Zhang, Baqun and Tsiatis, Anastasios A. and Davidian, Marie and Zhang, Min and Laber, Eric},
	year = {2012},
	pages = {103--114}
}

@incollection{robins2004optimal,
	address = {New York, NY},
	title = {Optimal {Structural} {Nested} {Models} for {Optimal} {Sequential} {Decisions}},
	isbn = {978-1-4419-9076-1},
	url = {https://doi.org/10.1007/978-1-4419-9076-1_11},
	doi = {10.1007/978-1-4419-9076-1_11},
	language = {en},
	urldate = {2026-02-21},
	booktitle = {Proceedings of the {Second} {Seattle} {Symposium} in {Biostatistics}: {Analysis} of {Correlated} {Data}},
	publisher = {Springer},
	author = {Robins, James M.},
	editor = {Lin, D. Y. and Heagerty, P. J.},
	year = {2004},
	pages = {189--326}
}

@article{murphy_experimental_2005,
	title = {An experimental design for the development of adaptive treatment strategies},
	volume = {24},
	copyright = {Copyright © 2004 John Wiley \& Sons, Ltd.},
	issn = {1097-0258},
	url = {https://onlinelibrary.wiley.com/doi/abs/10.1002/sim.2022},
	doi = {10.1002/sim.2022},
	language = {en},
	number = {10},
	urldate = {2026-02-22},
	journal = {Statistics in Medicine},
	author = {Murphy, S. A.},
	year = {2005},
	note = {\_eprint: https://onlinelibrary.wiley.com/doi/pdf/10.1002/sim.2022},
	pages = {1455--1481}
}

@article{watkins_technical_1992,
	title = {Technical {Note}: {Q}-{Learning}},
	volume = {8},
	issn = {1573-0565},
	shorttitle = {Technical {Note}},
	url = {https://doi.org/10.1023/A:1022676722315},
	doi = {10.1023/A:1022676722315},
	language = {en},
	number = {3},
	urldate = {2026-02-22},
	journal = {Machine Learning},
	author = {Watkins, Christopher J.C.H. and Dayan, Peter},
	month = may,
	year = {1992},
	pages = {279--292},
}

@book{vanderlaan2011targeted,
  title={Targeted learning: causal inference for observational and experimental data},
  author={{van der Laan}, Mark J and Rose, Sherri},
  year={2011},
  edition={1st},
  publisher={Springer New York},
  doi={10.1007/978-1-4419-9782-1}
}

@article{luedtke2016super,
	title = {Super-{Learning} of an {Optimal} {Dynamic} {Treatment} {Rule}},
	volume = {12},
	issn = {1557-4679},
	url = {https://www.degruyterbrill.com/document/doi/10.1515/ijb-2015-0052/html},
	doi = {10.1515/ijb-2015-0052},
	language = {en},
	number = {1},
	urldate = {2026-02-22},
	journal = {The International Journal of Biostatistics},
	publisher = {De Gruyter},
	author = {Luedtke, Alexander R. and {van der Laan}, Mark J},
	month = may,
	year = {2016},
	pages = {305--332}
}

@misc{suk2024fair,
	title = {Fair and {Robust} {Estimation} of {Heterogeneous} {Treatment} {Effects} for {Optimal} {Policies} in {Multilevel} {Studies}},
	url = {https://osf.io/preprints/psyarxiv/xz3jw_v1/},
	doi = {10.31234/osf.io/xz3jw},
	urldate = {2026-01-06},
	publisher = {PsyArXiv},
	author = {Suk, Youmi and Park, Chan and Pan, Chenguang and Kim, Kwangho},
	month = oct,
	year = {2024}
}

@article{suk_psychometric_2024,
	title = {A {Psychometric} {Framework} for {Evaluating} {Fairness} in {Algorithmic} {Decision} {Making}: {Differential} {Algorithmic} {Functioning}},
	volume = {49},
	issn = {1076-9986},
	shorttitle = {A {Psychometric} {Framework} for {Evaluating} {Fairness} in {Algorithmic} {Decision} {Making}},
	url = {https://doi.org/10.3102/10769986231171711},
	doi = {10.3102/10769986231171711},
	language = {EN},
	number = {2},
	urldate = {2026-02-25},
	journal = {Journal of Educational and Behavioral Statistics},
	publisher = {American Educational Research Association},
	author = {Suk, Youmi and Han, Kyung T.},
	month = apr,
	year = {2024},
	pages = {151--172}
}

@article{walkington2020appraising,
  title={Appraising research on personalized learning: Definitions, theoretical alignment, advancements, and future directions},
  author={Walkington, Candace and Bernacki, Matthew L},
  journal={Journal of Research on Technology in Education},
  volume={52},
  number={3},
  pages={235--252},
  year={2020},
  doi={10.1080/15391523.2020.1747757}
}

@book{xiong2022artificial,
  title={Artificial intelligence and causal inference},
  author={Xiong, Momiao},
  year={2022},
  publisher={Chapman and Hall/CRC},
  doi={10.1201/9781003028543}
}

@article{Robins1994Estimation,
author = {James M. Robins and Andrea Rotnitzky and Lue Ping Zhao},
title = {Estimation of Regression Coefficients When Some Regressors are not Always Observed},
journal = {Journal of the American Statistical Association},
volume = {89},
number = {427},
pages = {846--866},
year = {1994},
publisher = {Taylor \& Francis},
doi = {10.1080/01621459.1994.10476818}
}

@article{van2014target,
	title = {Targeted {Learning} of an {Optimal} {Dynamic} {Treatment}, and {Statistical} {Inference} for its {Mean} {Outcome}},
	url = {https://biostats.bepress.com/ucbbiostat/paper329},
	journal = {U.C. Berkeley Division of Biostatistics Working Paper Series},
	author = {{van der Laan}, Mark J and Luedtke, Alexander},
	month = sep,
	year = {2014}
}

@article{van2015target,
	title = {Targeted {Learning} of the {Mean} {Outcome} under an {Optimal} {Dynamic} {Treatment} {Rule}},
	volume = {3},
	issn = {2193-3685, 2193-3677},
	url = {https://www.degruyterbrill.com/document/doi/10.1515/jci-2013-0022/html},
	doi = {10.1515/jci-2013-0022},
	language = {en},
	number = {1},
	urldate = {2025-12-15},
	journal = {Journal of Causal Inference},
	author = {{van der Laan}, Mark J and Luedtke, Alexander R},
	month = mar,
	year = {2015},
	pages = {61--95}
}

@article{white2011multiple,
	title = {Multiple imputation using chained equations: {Issues} and guidance for practice},
	volume = {30},
	copyright = {Copyright © 2010 John Wiley \& Sons, Ltd.},
	issn = {1097-0258},
	url = {https://onlinelibrary.wiley.com/doi/abs/10.1002/sim.4067},
	doi = {10.1002/sim.4067},
	language = {en},
	number = {4},
	urldate = {2026-01-17},
	journal = {Statistics in Medicine},
	author = {White, Ian R. and Royston, Patrick and Wood, Angela M.},
	year = {2011},
	pages = {377--399}
}

@misc{pan2026learning,
	title = {Learning {Feasible} {Optimal} {Treatment} {Regimes} for {Personalized} {Decision}-{Making}},
	url = {https://osf.io/preprints/psyarxiv/arp48_v1/},
	doi = {10.31234/osf.io/arp48_v1},
	urldate = {2026-01-06},
	publisher = {PsyArXiv},
	author = {Pan, Chenguang and Li, Yuxuan and Suk, Youmi},
	month = jan,
	year = {2026},
}

@software{malenica2022tmle3mopttx,
      author = {Malenica, Ivana and Coyle, Jeremy and {van der Laan}, Mark J},
      title = {{tmle3mopttx}: Targeted Learning and Variable Importance with Optimal Individualized Categorical Treatment},
      year  = {2022},
      doi = {},
      url = {https://github.com/tlverse/tmle3mopttx},
      note = {R package version 1.0.0}
    }

@software{Jeremy2023sl3,
  title = {{sl3}: Pipelines for Machine Learning and {Super Learning}},
  author = {Coyle, Jeremy and Hejazi, Nima and Malenica, Ivana and Phillips, Rachael and Sofrygin, Oleg}, % All Last, First
  year = {2023},
  note = {R package version 1.4.5},
  doi = {10.5281/zenodo.1342293},
  url = {https://github.com/tlverse/sl3},
}

@software{jeremy2021tmle3,
  title = {{tmle3}: The Extensible {TMLE} Framework},
  author = {Jeremy Coyle},
  year = {2021},
  note = {R package version 0.2.0},
  doi = {10.5281/zenodo.4603358},
  url = {https://github.com/tlverse/tmle3},
}

@Manual{R2025,
  title = {R: A Language and Environment for Statistical Computing},
  author = {{R Core Team}},
  organization = {R Foundation for Statistical Computing},
  address = {Vienna, Austria},
  year = {2025},
  url = {https://www.R-project.org/},
}

@incollection{zheng2011cvtmle,
  title     = {Cross-Validated Targeted Minimum-Loss-Based Estimation},
  author    = {Zheng, Wenjing and {van der Laan}, Mark J},
  booktitle = {Targeted Learning: Causal Inference for Observational and Experimental Data},
  publisher = {Springer},
  address   = {New York, NY},
  year      = {2011},
  pages     = {459--474},
  doi       = {10.1007/978-1-4419-9782-1_27}
}

@article{vanderlaan2006tml,
  title   = {Targeted Maximum Likelihood Learning},
  author  = {{van der Laan}, Mark J and Rubin, Daniel},
  journal = {The International Journal of Biostatistics},
  year    = {2006},
  volume  = {2},
  number  = {1},
  doi     = {10.2202/1557-4679.1043}
}

@article{rubin1974estimating,
  title={Estimating causal effects of treatments in randomized and nonrandomized studies},
  author={Rubin, Donald B},
  journal={Journal of educational Psychology},
  volume={66},
  number={5},
  pages={688},
  year={1974},
  publisher={American Psychological Association},
  doi={10.1037/h0037350}
}

@article{rubin1986comment,
  title={Comment: Which ifs have causal answers},
  author={Rubin, Donald B},
  journal={Journal of the American statistical association},
  volume={81},
  number={396},
  pages={961--962},
  year={1986},
  publisher={Taylor \& Francis}
}

@book{imbens2015causal,
  title={Causal inference in statistics, social, and biomedical sciences},
  author={Imbens, Guido W and Rubin, Donald B},
  year={2015},
  publisher={Cambridge university press}
}

@article{apley2020visualizing,
  author    = { Apley, Daniel W. and Zhu, Jingyu },
  title     = {{Visualizing the Effects of Predictor Variables in Black Box Supervised Learning Models}},
  journal   = {Journal of the Royal Statistical Society Series B: Statistical Methodology},
  year      = {2020},
  volume    = {82},
  number    = {4},
  pages     = {1059--1086},
  doi       = {10.1111/rssb.12377},
}

@article{hooker2004discovering,
  author    = { Hooker, Giles },
  title     = {{Discovering additive structure in black box functions}},
  journal   = {Proceedings of the tenth ACM SIGKDD international conference on Knowledge discovery and data mining},
  year      = {2004},
  pages     = {575--580},
  doi       = {10.1145/1014052.1014122},
}

@article{hooker2007generalized,
  author    = { Hooker, Giles },
  title     = {{Generalized Functional ANOVA Diagnostics for High-Dimensional Functions of Dependent Variables}},
  journal   = {Journal of Computational and Graphical Statistics},
  year      = {2007},
  volume    = {16},
  number    = {3},
  pages     = {709--732},
  doi       = {10.1198/106186007x237892},
}

@inproceedings{lundberg2017unified,
  author    = { Lundberg, Scott and Lee, Su-In },
  title     = {{A Unified Approach to Interpreting Model Predictions}},
  booktitle = {Advances in Neural Information Processing Systems 30 (NIPS 2017)},
  year      = {2017},
  pages     = {4765--4774},
  doi       = {10.48550/arXiv.1705.07874},
  url       = {https://arxiv.org/abs/1705.07874},
}

@article{simpkins2006math,
	address = {US},
	title = {Math and science motivation: {A} longitudinal examination of the links between choices and beliefs},
	volume = {42},
	issn = {1939-0599},
	shorttitle = {Math and science motivation},
	doi = {10.1037/0012-1649.42.1.70},
	number = {1},
	journal = {Developmental Psychology},
	publisher = {American Psychological Association},
	author = {Simpkins, Sandra D. and Davis-Kean, Pamela E. and Eccles, Jacquelynne S.},
	year = {2006},
	pages = {70--83}
}

@article{mceachin2020heterogeneous,
	title = {Heterogeneous {Effects} of {Early} {Algebra} across {California} {Middle} {Schools}},
	volume = {39},
	issn = {1520-6688},
	url = {https://onlinelibrary.wiley.com/doi/abs/10.1002/pam.22202},
	doi = {10.1002/pam.22202},
	language = {en},
	number = {3},
	urldate = {2026-08-08},
	journal = {Journal of Policy Analysis and Management},
	author = {McEachin, Andrew and Domina, Thurston and Penner, Andrew},
	year = {2020},
	pages = {772--800}
}

@incollection{bohrnstedt2020mathematics,
	title = {Mathematics {Identity}, {Self}-efficacy, and {Interest} and {Their} {Relationships} to {Mathematics} {Achievement}: {A} {Longitudinal} {Analysis}},
	isbn = {978-3-030-41231-9},
	url = {https://doi.org/10.1007/978-3-030-41231-9_7},
	doi = {10.1007/978-3-030-41231-9_7},
	language = {en},
	urldate = {2026-08-14},
	booktitle = {Identity and {Symbolic} {Interaction}: {Deepening} {Foundations}, {Building} {Bridges}},
	publisher = {Springer International Publishing},
	author = {Bohrnstedt, George W. and Zhang, Jizhi and Park, Bitnara Jasmine and Ikoma, Sakiko and Broer, Markus and Ogut, Burhan},
	editor = {Serpe, Richard T. and Stryker, Robin and Powell, Brian},
	year = {2020},
	pages = {169--210},
}

@article{mcdonald2019single,
	title = {A {Single}-{Item} {Measure} for {Assessing} {STEM} {Identity}},
	volume = {4},
	issn = {2504-284X},
	url = {https://www.frontiersin.org/journals/education/articles/10.3389/feduc.2019.00078/full},
	doi = {10.3389/feduc.2019.00078},
	language = {English},
	urldate = {2026-08-14},
	journal = {Frontiers in Education},
	publisher = {Frontiers},
	author = {McDonald, Melissa M. and Zeigler-Hill, Virgil and Vrabel, Jennifer K. and Escobar, Martha},
	month = jul,
	year = {2019}
}
